\documentclass[fleqn,usenatbib]{mnras}
\usepackage{newtxtext,newtxmath}

\usepackage[T1]{fontenc}

\DeclareRobustCommand{\VAN}[3]{#2}
\let\VANthebibliography\thebibliography
\def\thebibliography{\DeclareRobustCommand{\VAN}[3]{##3}\VANthebibliography}

\usepackage{graphicx}	
\usepackage{amsmath}	
\usepackage{array}
\usepackage{float}

\usepackage{subcaption}
\usepackage{stfloats}

\usepackage{multicol} 
\usepackage{xcolor, colortbl}

\newcolumntype{P}[1]{>{\centering\arraybackslash}p{#1}}
\newcolumntype{M}[1]{>{\centering\arraybackslash}m{#1}}

\usepackage{orcidlink}

\title[Testing GaiaX for EM Follow-Up of GW events
]{Preparing for Gaia Searches for Optical Counterparts of Gravitational Wave Events during O4}

\author[S. Biswas et al.]{Sumedha Biswas\orcidlink{0009-0006-7543-1544},$^{1}$\thanks{E-mail: s.biswas@astro.ru.nl}
Zuzanna Kostrzewa-Rutkowska,$^{3}$
Peter G. Jonker\orcidlink{0000-0001-5679-0695 }$^{1,2}$,
Paul Vreeswijk$^{1}$, 
\newauthor{Deepak Eappachen\orcidlink{0000-0001-7841-0294}$^{2,1}$, 
Paul J. Groot$^{1,4,5}$, 
Simon Hodgkin$^{6}$,
Abdullah Yoldas$^{6}$, 
Guy Rixon$^{6}$, 
Diana Harrison$^{6,7}$}, 
\newauthor{M. van Leeuwen$^{6}$, 
Dafydd Evans$^{6}$}
\\
$^{1}$Department of Astrophysics/IMAPP, Radboud University, PO Box 9010, 6500 GL Nijmegen, The Netherlands\\
$^{2}$SRON, Netherlands Institute for Space Research, Nie ls Bohrweg 4, 2333 CA Leiden, the Netherlands\\
$^{3}$Leiden Observatory, Leiden University, PO Box 9513, NL-2300 RA Leiden, the Netherlands\\
$^{4}$South African Astronomical Observatory, PO Box 9, Observatory, 7935, Cape Town, South Africa\\
$^{5}$Department of Astronomy, University of Cape Town, Private Bag X3, Rondebosch, 7701, South Africa\\
$^{6}$Institute of Astronomy, University of Cambridge, Madingley Road, Cambridge CB3 0HA, UK\\
$^{7}$Kavli Institute for Cosmology Cambridge, Institute of Astronomy, Madingley Road, Cambridge, CB3 0HA
}

\date{Accepted XXX. Received YYY; in original form ZZZ}

\pubyear{2023}

\begin{document}
\label{firstpage}
\pagerange{\pageref{firstpage}--\pageref{lastpage}}

\maketitle

\begin{abstract}
The discovery of gravitational wave (GW) events and the detection of electromagnetic counterparts from GW170817 has started the era of multimessenger GW astronomy. The field has been developing rapidly and in this paper, we discuss the preparation for detecting these events with ESA's {\it Gaia} satellite, during the 4th observing run of the LIGO-Virgo-KAGRA (LVK) collaboration that has started on May 24, 2023. \textit{Gaia} is contributing to the search for GW counterparts by a new transient detection pipeline called \textit{GaiaX}. In \textit{GaiaX}, a new source appearing in the field of view of only one of the two telescopes on-board {\it Gaia} is sufficient to send out an alert on the possible detection of a new transient. Ahead of O4, an experiment was conducted over a period of about two months. During the two weeks around New Moon in this period of time, the MeerLICHT (ML) telescope located in South Africa tried (weather permitting) to observe the same region of the sky as \textit{Gaia} within 10 minutes. Any \textit{GaiaX} detected transient was published publicly. ML and {\it Gaia} have similar limiting magnitudes for typical seeing conditions at ML. At the end of the experiment, we had 11861 \textit{GaiaX} candidate transients and 15806 ML candidate transients, which we further analysed and the results of which are presented in this paper. Finally, we discuss the possibility and capabilities of \textit{Gaia} contributing to the search for electromagnetic counterparts of gravitational wave events during O4 through the {\it GaiaX} detection and alert procedure.
\end{abstract}




\section{Introduction}

The discovery of gravitational waves (GW) from the merger of a binary black hole (BBH) during the first observing run (O1) of the Laser Interferometer Gravitational-Wave Observatory (LIGO)/Virgo collaboration in 2015 marked the beginning of gravitational wave astrophysics \citep{gw150914}. Subsequent observing runs (O2 and O3) have resulted in the detection of more than 50 BBH mergers \citep{gwtc1, gwtc2}. O2 yielded a significant discovery in the form of the binary neutron star (BNS) merger event GW170817 \citep{Abbott_2017}, which is currently the sole instance of a BNS merger detected with an accompanying electromagnetic (EM) counterpart. This landmark detection marks the advent of the multi-messenger era in astrophysics. The EM counterpart, known as AT2017gfo, was a kilonova observed at a distance of 40~Mpc and was detected across the EM spectrum, from radio to $\gamma$-rays \citep{2017Andreoni,2017Arcavi,2017Chornock,2017Coulter,2017Covino,2017Cow,2017Drout,2017Kasliwal,2017Evans,2017Lipunov,2017Nicholl,2017Pian,2017Shappee,2017Smartt,2017Tanvir,2017Troja,Yang_2019,2017utsumi, Soares-Santos_2017}.

Compact Binary Coalescence (CBC) events involving at least one neutron star (NS), i.e. BNS and NSBH binaries, can result in multiple associated EM signals, e.g., a beamed gamma-ray burst (GRB) and its afterglow, and a kilonova \citep{Metzger_2012}. GRBs are highly collimated and relativistic outflows \citep{Berger_2014} that are thought to be driven by the accretion of a massive remnant disk onto the massive NS, or BH, remnant following the CBC merger (e.g., \citet{1992narayan}). A GRB associated with a GW event is expected to occur within a couple of seconds after the merger. Kilonovae are thermal transients lasting days to weeks \citep{Smartt_2017}, and are powered by the radioactive decay of heavy, neutron-rich, elements synthesised through the $r$-process in the expanding merger ejecta \citep{optical_counterparts_1998, Tanvir_2017}. It has been suggested that a Fast X-ray Transient (FXT) signal might be generated in the immediate aftermath of a BNS merger if the merger product is a massive, rapidly spinning magnetar \citep{metzgerpiro, Sun_2017, Sun_2019, Metzger_2017}. Such FXT signals have been detected, although currently the link with BNS mergers is tenuous \citep{Jonker_2013, 2017Bauer, 2019lin, 2021lin, Quirola-Vasquez:2022nea, Quirola-Vasquez:2023eye, Eappachen:2023wrs, metzgerpiro, Sun_2017, Sun_2019, Zheng_2017}. During O3, two binary neutron star-black hole (NSBH) mergers \citep{lvk_nsbh} were detected, however, for neither an EM counterpart was detected. During the current observing run O4, that started on May 24, 2023, the predicted rate for a BNS event, from which an EM counterpart is most probable, is approximately 1 event every 3 weeks \citep{2023lscpop}. The scientific yield of finding EM signals coincident with a GW event is large, including for instance contributions to determining the neutron star equation-of-state (e.g., \citealt{Lattimer_2012, De_2018_EoS, Guerra_Chaves_2019}), understanding $r$-process nucleosynthesis (e.g., \citealt{Metzger_2017, 2017Villar, Hotokezaka_2018}), and measuring the Hubble constant (e.g., \citealt{gwhubbleconstant2017, bulla2022multimessenger}). As a result, the hunt for EM counterparts is a rapidly developing field, with its own set of challenges.

One of the primary difficulties encountered when conducting a follow-up campaign of a GW event is the necessity for low latency and sensitive observations, combined with the need to examine large sky localization areas of hundreds of square degrees. In the aftermath of a CBC event detection, such as the merger of a BNS or NSBH system, the optical emission is predicted to be detectable for a short period of time only, necessitating prompt follow-up. The duration of the optical counterpart's detectability during O4 is challenging to predict precisely, however, previous observations of AT2017gfo suggest that it could range from a few hours to several days to at best a week for most optical survey telescopes (see e.g., figure~1 in \citealt{2017Villar}). The probability of detecting the optical counterpart depends on the luminosity distance of the CBC event; with the O4 detector sensitivity, events at a distance of up to 190~Mpc can be detected (\href{https://dcc.ligo.org/LIGO-G2002127/public}{LIGO-G2002127-v18}). It also depends on the number of online and locked GW detectors, as well as the signal-to-noise ratio (SNR) of the GW signal, and both factors have a significant impact on the sky localization area for a GW event. It can extend up to hundreds or thousands of square degrees across the sky \citep{Abbott_2020_locali, Gehrels_2016}. Altogether, the follow-up of GW events is a challenging endeavour and over the years, it has been tackled in different ways \citep{Ghosh_2016, Salafia_2017, Rana_2017, Chan__2017, Coughlin_2016}. 

 
The European Space Agency (ESA) \textit{Gaia} satellite has been operational since mid-2014 and it is positioned at the second Lagrange (L2) point of the Sun-Earth-Moon system \citep{gaia}. The latest release of \textit{Gaia} data, known as Gaia Data Release 3 (GDR3), comprises accurate astrometric and photometric measurements for over 1.8 billion sources that are brighter than magnitude 21 \citep{gaia_dr3}. In order to make these measurements, \textit{Gaia} scans the full sky. Each observation consists of a 50~s long white-light (G-band) light curve, sampled every 5~s, which can in principle also be used for variability detection on very short timescales \citep{Wevers_2018, Roelens_2018}. During the nominal 5-year mission duration, \textit{Gaia} scanned over each sky location from different angles more than 70 times. During the mission extension phase, these repeat visits continue. This enables \textit{Gaia} to detect transients, which are published in a public alerts stream, known as the Gaia Science Alerts (GSA) \citep{Hodgkin_gsa}. During O3, the GSA were examined to identify any potential matches between the location and times of the GSA-transients and the sky localisation regions and times of GW detections \citep{190408gaia, GCN_190426, GCN_190425,  190510gaia, 190517gaia, 190503gaia, 190521gaia, 190630gaia, 190718gaia, 190901gaia, 190910gaia, 191205gaia}.


MeerLICHT (ML) and BlackGEM (BG) are both wide-field and fully robotic telescopes, built to study transient phenomena and develop a southern sky survey at declination angles $<30^\circ$ in the 6 optical filters that ML and BG employ. All four (one for ML, three for BG) telescopes have a 65~cm primary mirror and are equipped with a single 10560~x~10560 pixel STA1600 detector which provides a wide field-of-view of 2.7~${deg}^2$ at 0.56$\arcsec$/pixel \citep{MLBG_SPIE}. The BG array situated at ESO La Silla (Chile), is designed for the optical follow-up of GW events. BG is capable of deeper observations than ML due to the better seeing conditions at La Silla. For this work, however, we used the MeerLICHT (ML) telescope, which is the prototype of BG situated at the South African Astronomical Observatory (SAAO) near Sutherland, as BG was not yet operational at the time of our observations reported here. 


Over the 2-week period around New Moon during September and October 2021, we ran an experiment where ML and {\it Gaia} observed the same region of the sky within 10 minutes, subject to the atmospheric conditions at Sutherland. During the whole $\sim2$ months period, the \textit{GaiaX} alerts were published (see: \url{https://www.cosmos.esa.int/web/gaia/iow_20210825}) and we compared the alerts to the transient candidates detected by ML if the {\it GaiaX} alert and the candidate ML transient were from a source that was observed within 10 minutes by both facilities. The objective for this experiment was multi-fold: 

\begin{enumerate}
    \item To test and improve the filtering applied to the \textit{GaiaX} alerts stream tailored to remove as many false positives as possible, before turning it on during O4. 
    \item To test the transient detection pipelines and help improve the ML/BG  bogus filtering techniques and machine-learning algorithms that classify candidate transients as real or bogus that are in place.
    \item To investigate any interesting real transient candidate that we might find.
\end{enumerate}

This paper is organized as follows: in Section~\ref{sec:gaia}, we describe the \textit{Gaia} telescope and in Section~\ref{sec:gaiax}, we discuss the \textit{GaiaX} detection pipeline; we describe the ML telescope in Section~\ref{sec:mlbg} and the ML/BG transient detection pipeline in Section~\ref{sec:mlbg_software}; in Section~\ref{sec:results}, we discuss the experiment and the results in detail; finally, we conclude with Section~\ref{sec:remarks}.

\section{Gaia}
\label{sec:gaia}
The ESA-{\it Gaia} spacecraft is equipped with two identical telescopes, with apertures of 1.45~m $\times$ 0.50~m pointing in directions separated by the basic angle ($\Gamma$~=~106.5$^{\circ}$) \citep{gaia}. With a focal plane containing more than 100 CCD detectors, {\it Gaia} observes $\sim$1000 square degrees daily, down to a magnitude of G$\sim$20.7~mag. The spacecraft scans the sky in accordance with a scanning law described in \cite{gaia}, which prescribes the intended spacecraft pointing as a function of time.

\subsection{GaiaX Detection Algorithm}
\label{sec:gaiax}
The Data Processing and Analysis Consortium (DPAC) handles {\it Gaia}'s data flow, and ensures the publication of detected transients typically within 24-48 hours of observation through the GSA stream \citep{Hodgkin_gsa}. With an improved detection algorithm, the GSA can capture transient events that the existing detection algorithm might miss since they are too faint and/or too fast. Specifically, a bespoke detection algorithm has been introduced during the current observing run (O4) of the LIGO/Virgo/KAGRA (LVK) GW detectors. This detection algorithm, termed \textit{GaiaX}, is capable of detecting fainter transients \citep{Kostrzewa_Rutkowska_2020}, enhancing the possibility of \textit{Gaia}'s contribution to the search for EM counterparts. \textit{GaiaX} will run independent of the GW event triggers, but the alerts can coincide with a GW event sky localization area during O4. The details of this algorithm are discussed in depth in \cite{Kostrzewa_Rutkowska_2020}, however, we repeat certain key features in this section.

\begin{enumerate}
    \item \textbf{One Telescope Requirement:} The existing GSA detection algorithm triggers a detection if an event has been detected by both of \textit{Gaia}'s telescopes. This requirement is reduced in the \textit{GaiaX} pipeline to an observation in either telescope, thereby increasing the rate of finding transient events. Unfortunately, this also implies that the algorithm is more susceptible to the detection of different artefacts (for instance, asteroids) causing the rate of false-positive detections to increase, if no counter-measures are taken.
    
    \item \textbf{Magnitude Limit Increased:} The {\it GaiaX} pipeline requires the median 9(8) CCD flux to be $>$101.25~$e^{-}$/s, which is equivalent to \textit{G}~$\sim$~20.68~mag, calibrated as in the GSA pipeline. This is greater than the GSA limit of \textit{G}~$\sim$~19 mag or 475.77~$e^{-}$/s, thereby allowing the detection of fainter transients.
    
    \item \textbf{Known Sources and Dense Regions in the Sky Considered:} The \textit{GaiaX} pipeline makes use of all the \textit{Gaia} photometric data collected and published in the {\it Gaia} Data Releases (\citealt{brown_dr1, gdr2, gaia_dr3}) to identify new transient events. By employing this approach, it becomes feasible to detect any previously known source within the GW skymap area, thereby facilitating the identification of any new transient event more easily. If a candidate transient is within 0.5$\arcsec$ of a previously detected source in the GSA database, it will be considered as the same source and will not trigger a new alert.
    
    Additionally, artefacts introduced by bright stars, close binary systems, planets, solar system objects (SSOs) and known asteroids have all been removed, to help the filtering of false-positives. Additionally, candidate transient alerts that occur in dense regions like the Milky Way bulge and disc, the Large and Small Magellanic Clouds are filtered out and not reported (see figure~2 in \citealt{Kostrzewa_Rutkowska_2020}).
    
    \item \textbf{Sources Created During Astrometric Excursions of the Satellite}: Occasionally, during astrometric excursions of the satellite caused by, for instance, hits by minor meteorites and space debris (see \citealt{vanleeuwen_2007}), that temporarily bring the spacecraft off its course, artificially and erroneously new candidate transients are reported by the system (this happens because temporarily the spacecraft system does not accurately know its position). The number of observed sources as a function of time was considered to identify peaks during which the rate of detection of candidate transients is very high. In the \textit{GaiaX} algorithm, the mean number of observed candidates in previous 5 runs is calculated and the windows of time where the number of observed sources is 10$\sigma$ larger than the average were removed to eliminate new sources created by these astrometric excursions (similar approach as described in \citealt{Kostrzewa_Rutkowska_2020}).
\end{enumerate}

\section{MeerLICHT/BlackGEM}
\label{sec:mlbg}
ML was installed at the South African Astronomical Observatory (SAAO) site near Sutherland in South Africa in 2017 \citep{MLBG_SPIE}. The filter wheel consists of 5 SDSS filters \textit{u, g, r, i, z} plus the wider $q$ band (covering 440-720 nm, nearly $g+r$). This makes it well-suited for multi-colour studies of the transient sky. Designed and built as a prototype for BG, the ML telescope follows the 64-antennae MeerKAT radio array. As a result, ML provides night-time optical coverage of the radio sky observed by MeerKAT in the wide $q$-band filter. 



\begin{figure*}
\minipage{\textwidth}
  \includegraphics[scale=0.75]{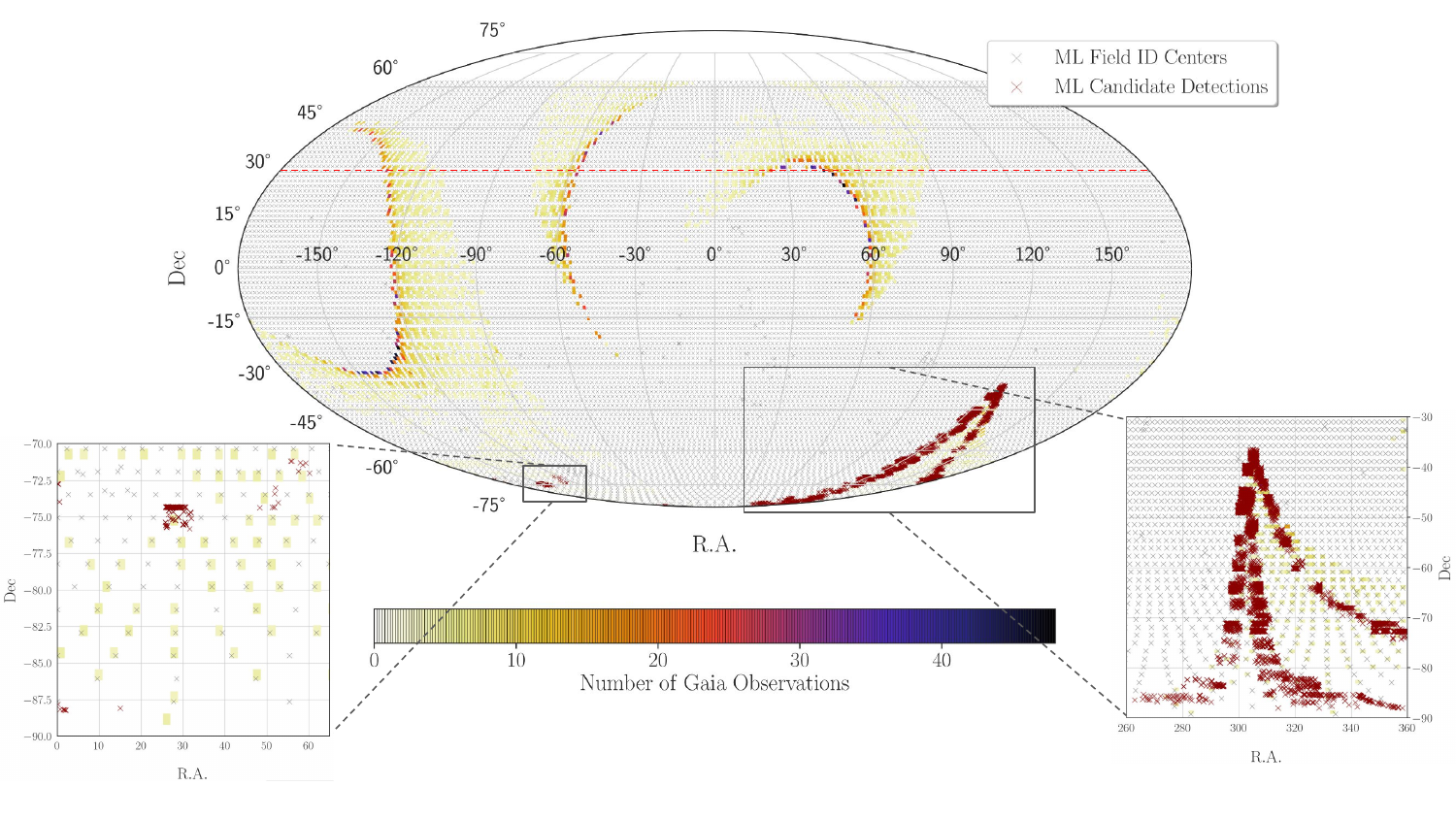}
\endminipage
\caption{The all-sky map of sky regions observed by {\it Gaia} and ML during the experiment, along with the ML field centres. The coloured density histogram shows the regions of the sky that were observed by Gaia, calculated with respect to the regions of the sky that ML can observe (ML field centres shown in grey) using the GOST tool. The colourbar indicates the number of times {\it Gaia} observed the same region of the sky over the 2 months of the experiment, with dark blue being the highest number of times and light yellow being the lowest number of times. With our set conditions of SNR\_ZOGY~$\leq$~12 and class\_real~$\geq$~0.7, ML had 15806 candidate transient detections during the period of the experiment which are shown by the dark red crosses. In principle, ML can observe at fields up to a declination of +60$^{\circ}$, however, the usual limit is at +30$^{\circ}$ (denoted by the red dashed line). The insets 1 and 2 show the zoomed-in ML candidate transient detections, where yellow is the region Gaia observed.}
\label{fig:gostplot}
\end{figure*}

\subsection{MeerLICHT/BlackGEM Transient Detection Pipeline}
\label{sec:mlbg_software}
The software pipeline to reduce raw images was initially written for ML, and was largely based on SkyMapper \citep{Scalzo_2017}, but now stands as an independent pipeline that has also been extended to the BG array. The pipeline consists of two integral components: BlackBOX\footnote{\label{blackbox}Source code at \href{https://github.com/pmvreeswijk/BlackBOX}{https://github.com/pmvreeswijk/BlackBOX}.} which performs standard CCD reduction tasks on the raw science images; and ZOGY\footnote{\label{zogy}Source code at \href{https://github.com/pmvreeswijk/ZOGY}{https://github.com/pmvreeswijk/ZOGY}.} which is used to identify sources, perform astrometry and photometry, and to identify transients using the optimal image subtraction routines that were formulated in \citet{ZOGY}. 

\begin{figure*}
    \includegraphics[width=0.82\textwidth]{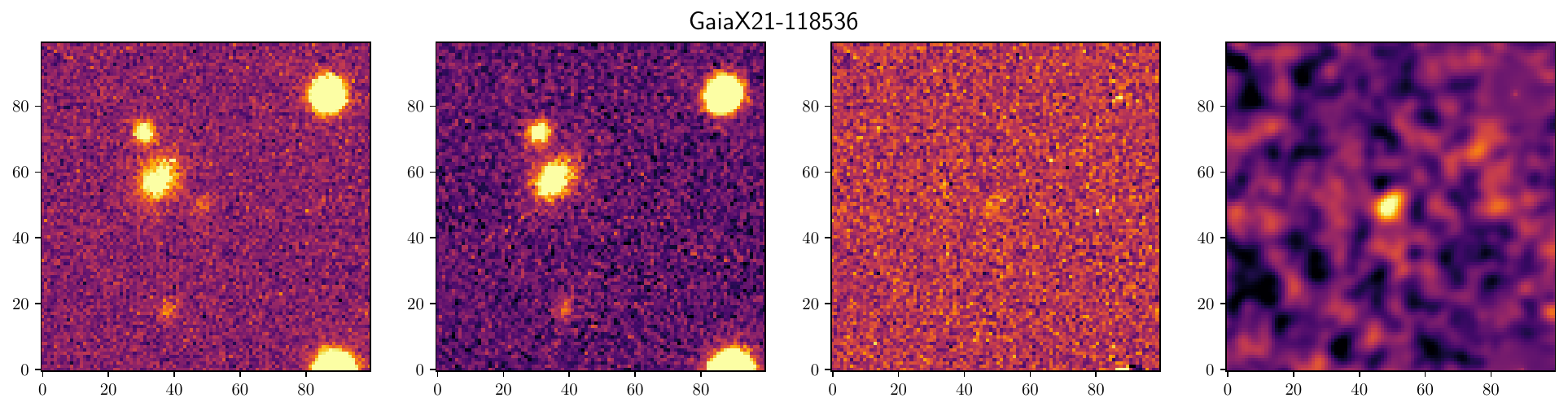}\\
    \includegraphics[width=0.82\textwidth]{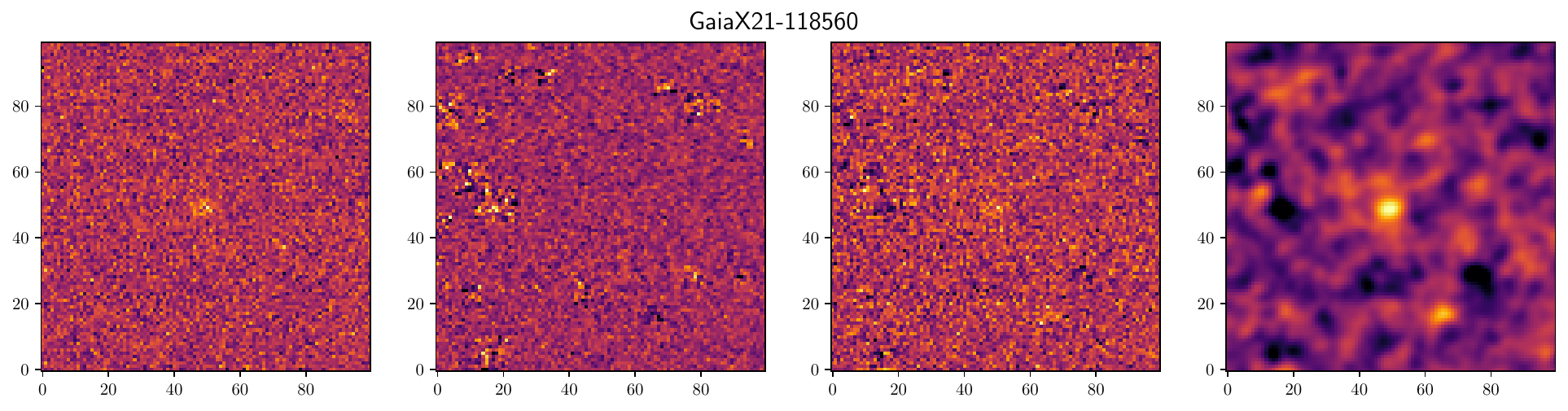}\\
    \includegraphics[width=0.82\textwidth]{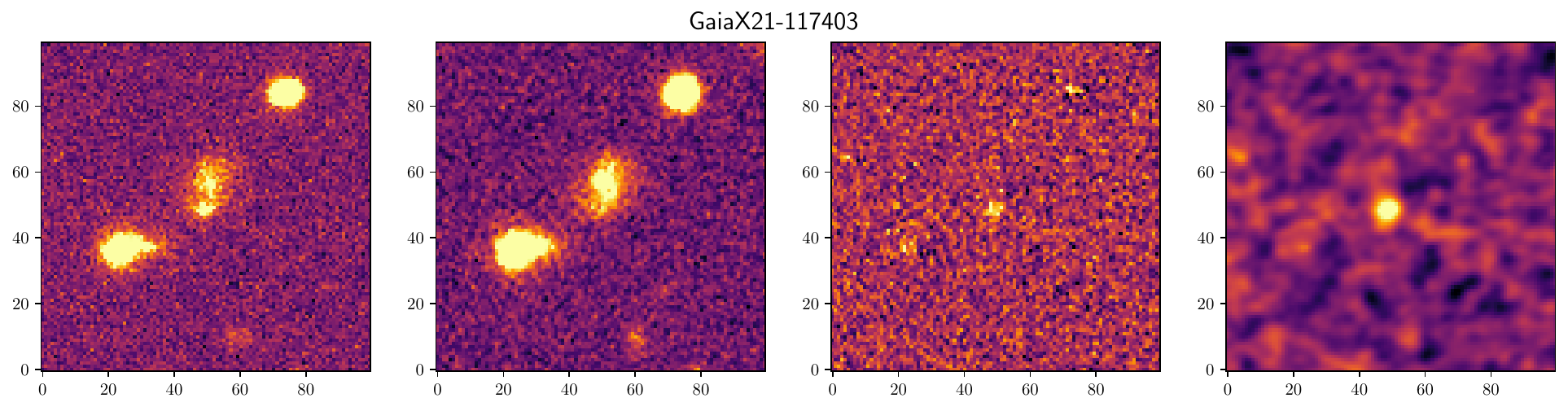}\\
    \includegraphics[width=0.82\textwidth]{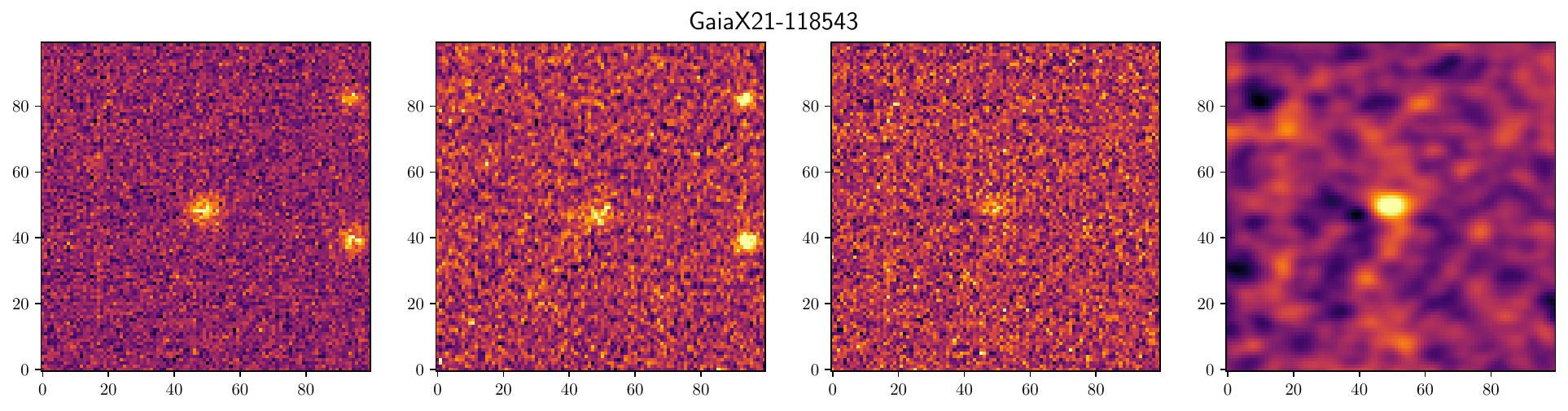}\\
    \includegraphics[width=0.82\textwidth]{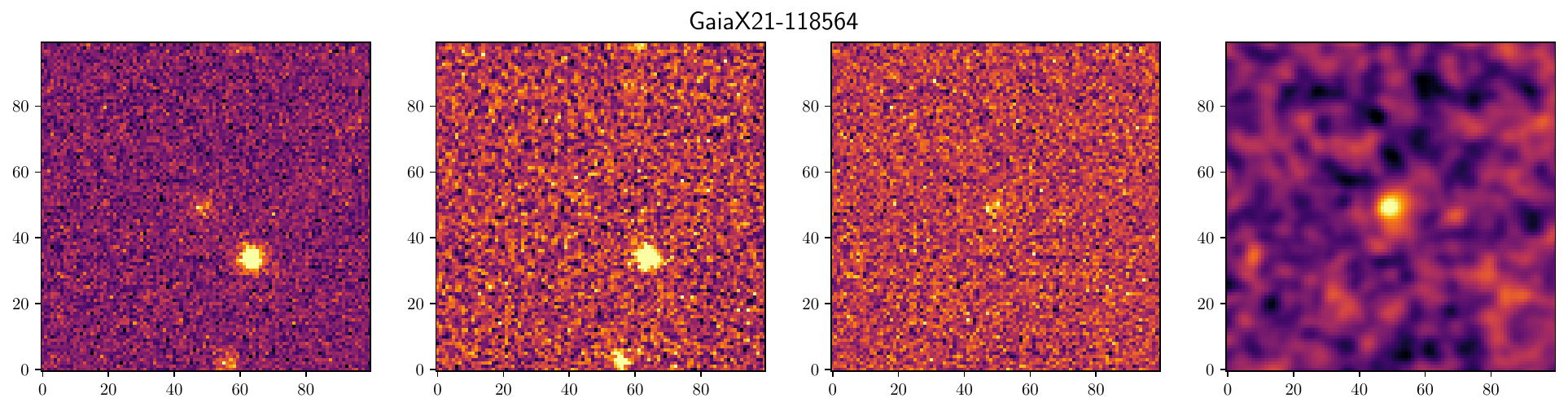}\\
    \includegraphics[width=0.82\textwidth]{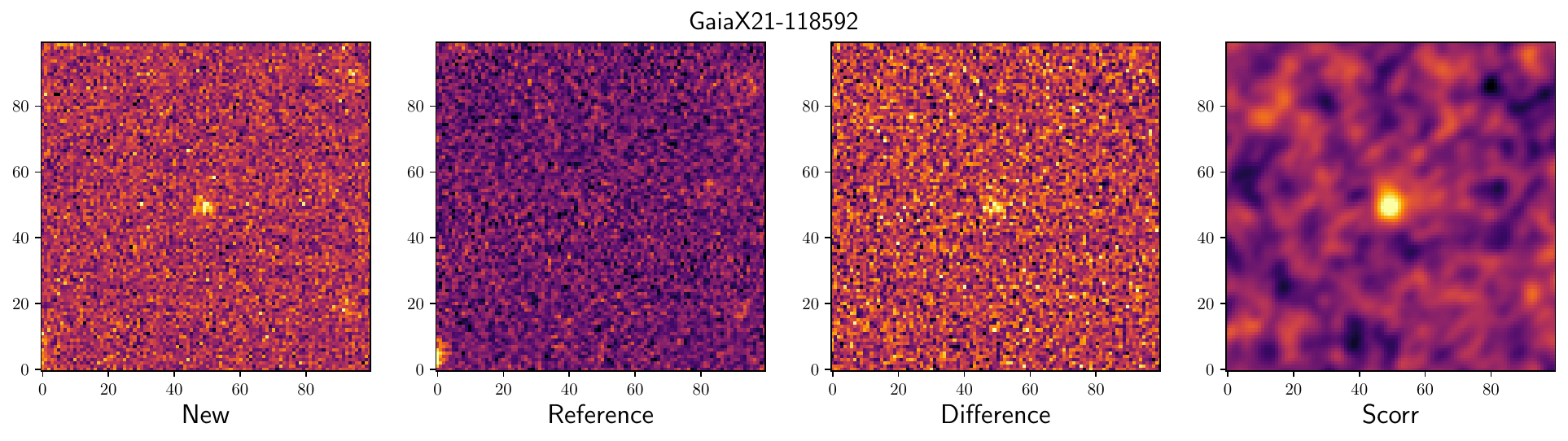}
    
    
    \caption{1\arcmin~$\times$~1\arcmin cut-outs of the transients centred on the transient position detected within ten minutes by ML and {\it GaiaX} (List \#1). {\it Left to Right:} the New, Reference, Difference and $S_{ \rm corr}$ images obtained from the ML database.
    }
\label{fig:real_transients}
\end{figure*}

The method uses statistical principles to derive the optimal statistic for transient detection, taking into account the point spread function of the new and reference image. The statistics image contains the probability of a transient being present at a particular location or pixel, while the corrected statistics ($S_{ \rm corr}$) or significance image is normalised using the source and background noise and astrometric uncertainties, resulting in an image having units of sigmas. The value of the $S_{ \rm corr}$ image at the position of an object is referred to as the ZOGY signal-to-noise ratio: \textit{SNR\_ZOGY}. Candidate transients are identified from the $S_{ \rm corr}$ images. All sources having a signal-to-noise ratio |SNR\_ZOGY| $\geq$ 6 are included in a transient catalogue file associated with the new image. A positive $S_{ \rm corr}$ value for a source indicates that the source is new or has brightened with respect to the reference image, while negative values indicate that the source has faded. The main image products of the transient detection pipeline is a set of 4 images - the reduced new, reduced reference, difference and $S_{ \rm corr}$ images (as seen in Figure~\ref{fig:real_transients}). 

In addition to these images, the transient detection pipeline also assigns a \textit{'real-bogus probability'} or \textit{'class-real'} value to each transient. This number can be anywhere between 0 and 1, with 1 being a real transient and 0 being a bogus (also known as false-positive) transient. This parameter is generated using machine-learning algorithms, and is described in further detail in \cite{meercrab}.

\begin{figure*}
\minipage{\textwidth}
\centering
  \includegraphics[width=\linewidth]{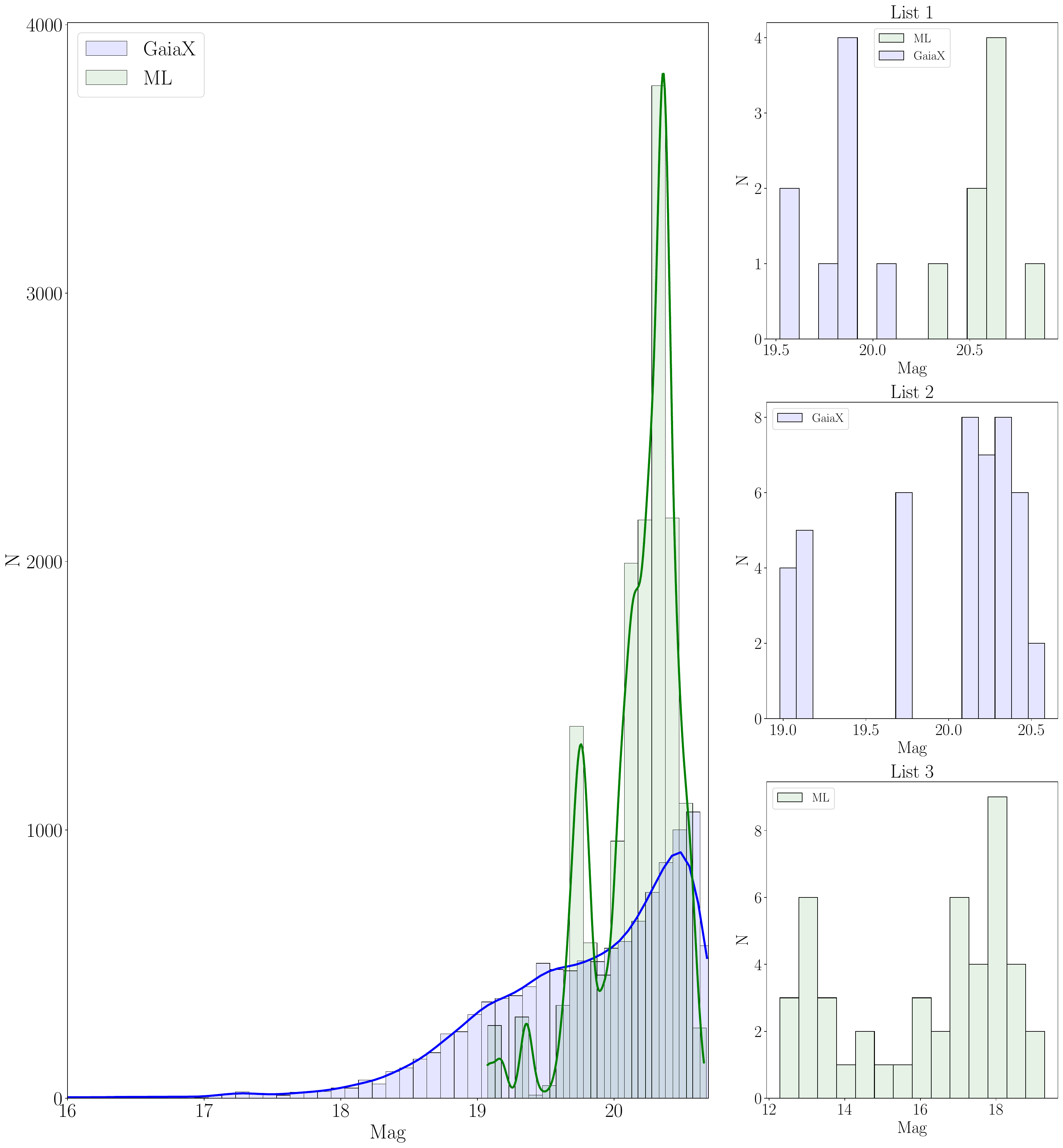}
\endminipage
\caption{{\it Left panel:} Number of candidate transients detected as a function of magnitude. The histogram shows the distribution for all the ML candidate transients (green) and all the \textit{GaiaX} alerts detected (blue) during the experiment period. The two solid lines indicate the kernel density estimation (KDE) \citep{kde1, kde2} of each of these distributions. {\it Right panels:}  Histograms of the number of candidate transients as a function of magnitude for each list described in Table~\ref{tab:lists}. {\it Top right panel:} List~\#1 refers to the 6 transients detected multiple times by both ML (green) and \textit{GaiaX} (blue). {\it Middle right panel:} List~\#2 consists of 21 \textit{GaiaX} alerts that were not detected by ML, shown in blue. {\it Bottom right panel:} List~\#3 in green, it refers to the 47 ML candidate transients that were not detected by \textit{GaiaX}.  }
\label{fig:allmag}
\end{figure*}

\section{Results and Discussion}
\label{sec:results}

During the period of two weeks centred on New Moon over about 2 months, 27th August 2021 to 4th November 2021, ML observed when visibilities were overlapping (and weather permitting) the same region of the sky as {\it Gaia} within 10 minutes. The \textit{GaiaX} Alerts pipeline was switched on for the whole period from August 27, 2021 to November 4, 2021 and the alerts were published. The position of {\it Gaia} on-sky as a function of time was determined using the {\it Gaia} Observation Forecast Tool (GOST)\footnote{\label{gost} \href{https://gaia.esac.esa.int/gost/}{https://gaia.esac.esa.int/gost/}.} which is based on the {\it Gaia} scanning laws. Using these positions as a function of time, ML pointed to the ML field centres such that the overlap with \textit{Gaia}'s scan was maximal. In Figure~\ref{fig:gostplot}, the GOST positions are plotted against the ML field centres to show the overlap in observations during the experiment. To determine the time of observations by ML, we used ObservationTimeAtGaia[UTC] provided by GOST, i.e. the observation time that {\it Gaia} looks in the chosen direction, here the ML field centre, expressed in Earth time. The only possible offsets are caused by the fact that the light observed in the ML telescope would be geometrically delayed or earlier due to the baseline offset of {\it Gaia} and Earth (up to 3 seconds), and negligible relativistic effects.


Once the experiment had concluded, \textit{GaiaX} had issued 11861 alerts between 2021-08-27 05:00:00 and 2021-11-04 21:07:00, out of which, there were 10535 unique candidate transients. During the same time, ML found 15806 candidate transient detections where we set the two parameters SNR\_ZOGY and class\_real of the ML transient detection pipeline to $\leq$~12 and $\geq$~0.7, respectively. Additionally, ML can, in principle, observe fields up to a declination of +40$^{\circ}$ but in practice the usual limit is set at +30$^{\circ}$ (airmass $<2$). Over this period and below this declination limit, there were 7572 unique \textit{GaiaX} alerts. We then checked the data to quantify how many nights ML was indeed observing the same region of the sky as \textit{Gaia}, and found that ML was observing the {\it Gaia} fields for 5 out of the 14 possible nights or for $\sim$35.7\% of the total time. While this number is small, it was not unexpected due to there being a small window every night when the visibilities align, and bad weather conditions at Sutherland on multiple nights reduced the number further. There were 251 ML and 74 \textit{GaiaX} candidate transients from when both telescopes were observing within ten minutes of each other. We then compared and analysed this reduced dataset, which is discussed below and summarized in Table~\ref{tab:lists} and detailed in Figure~\ref{fig:allmag}.

\begin{table*}
\caption{\label{tab:lists} Summary of the results of the experiment i.e. from the 5 days of contemporaneous observations by ML and \textit{Gaia}, divided into three lists with the corresponding number of ML candidate transients/{\it GaiaX} alerts for each.}
\centering
    \begin{tabular}{|M{0.7cm}|M{2.5cm}|M{2.5cm}|M{5.8cm}|M{1.1cm}|}
    \hline
    \textbf{List $\#$} & \textbf{GaiaX Alert?} & \textbf{MeerLICHT Transient?} & \textbf{Outcome} & \textbf{$\#$} \\
     \hline
     1 & \cellcolor{green!15}Yes & \cellcolor{green!15}Yes & Real Transient & 6  \\
    \hline
     2 & \cellcolor{green!15}Yes & \cellcolor{red!15}No & GaiaX detection is potentially spurious since the ML detection limit is deeper than the GaiaX detection magnitude of the transient & 21  \\
    \hline
    3 & \cellcolor{red!15}No & \cellcolor{green!15}Yes & The ML transient was too faint to be detected by GaiaX or the source was removed by the \textit{GaiaX} pipeline. & 47  \\
    \hline
    \end{tabular}
\end{table*}

\begin{figure}
    \includegraphics[width=\columnwidth, scale=1.5]{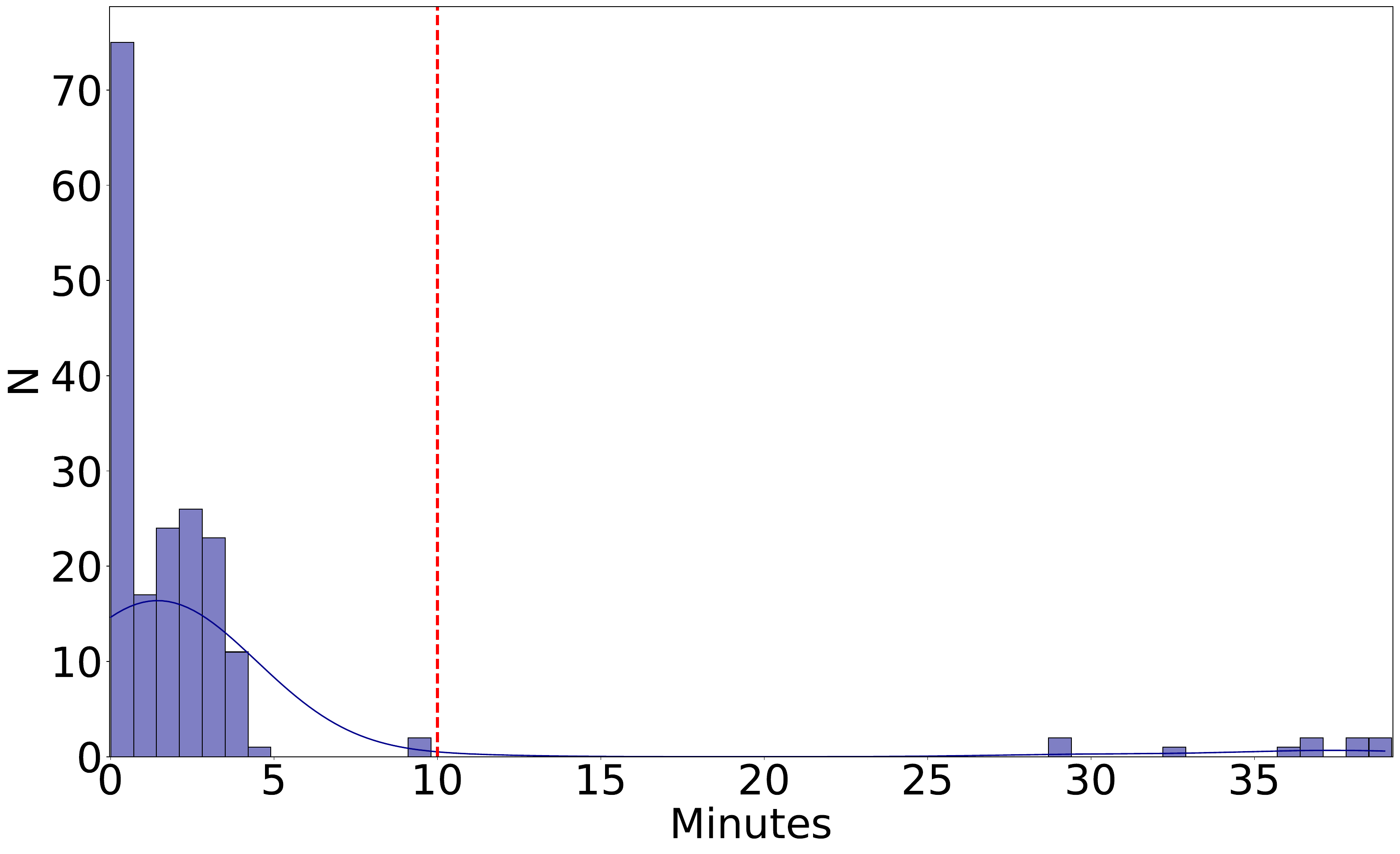}
    \caption{The time difference between the ML and {\it GaiaX} detections of candidate transients, indicating that majority of the detections (denoted by {\it N}) during the experiment were within 10 minutes of each other, as described in Section~\ref{sec:real}.}
    \label{fig:timediff}
\end{figure}

\subsection{List \#1: Real Transients}
\label{sec:real}
The first thing that we set out to determine is which of the transients are detected by both \textit{GaiaX} and ML; we consider these transients to be real. To do this, we applied the forced photometry routine developed for ML \footref{zogy} to the input list of \textit{GaiaX} transients, including their times of observation, to determine whether ML detected a significant transient at the same position around the same time. The forced photometry routine collects all relevant ML images based on their astrometric solution and their mid-exposure date of observation being within 10 minutes of the GaiaX time of observation (see Figure~\ref{fig:timediff}). From the relevant images it subsequently measures the magnitude, magnitude error, signal-to-noise ratio (SNR) and limiting magnitude at the GaiaX input positions; these quantities can optionally be measured from the reduced images, the reference images and/or the transient products. For the reduced and reference images (without any image subtraction), the magnitude measurements are made by weighing the source flux with the point spread function (PSF) at the source position; that position-dependent PSF is constructed for each ML image using the PSFEx package \citep{2011ASPC..442..435B}. This optimal magnitude determination closely follows the method described in \citet{1986PASP...98..609H} \citep[see also][]{1998MNRAS.296..339N}. Below, SNR\_OPT refers to the corresponding signal-to-noise ratio of a source in the reduced image. The transient (limiting) magnitudes are based on the PSF flux as determined by ZOGY (see Eq.~41 from \citet{ZOGY}), after the reference image has been optimally subtracted from the reduced image. Below, SNR\_ZOGY refers to the corresponding transient signal-to-noise ratio. The ML flux calibration will be described in detail in Vreeswijk et al.~(in prep.).

We detect 6 unique matching transient sources a total of 51 times when we require a ML transient to have {\rm SNR\_ZOGY}$>=3$, which are shown in Figure~\ref{fig:real_transients} and listed in Table~\ref{tab:list1}. These transients are also discussed below, and their light curve information is included in Table~\ref{tab:realtransient}.

\textit{GaiaX21-118536--} (RA=337.54984$^{\circ}$, Dec= --86.31725$^{\circ}$) On 2021-09-08 at 21:06:43 UTC, a source at a magnitude of 19.89~$\pm$~0.04 was detected by {\it GaiaX}. This source was also detected by ML on 2021-09-08 21:04:53 UTC at a magnitude of 20.02~$\pm$~0.16 in the $q$-band.

\textit{GaiaX21-118560--} (RA=307.08475$^{\circ}$, Dec=--74.79558$^{\circ}$) was detected as a \textit{GaiaX} Alert on 2021-09-08 23:06:34 UTC at a $G$-band magnitude of 20.13~$\pm$~0.02 and on 2021-09-08 23:03:23 UTC by ML at a magnitude of 20.11~$\pm$~0.18 in the $q$-band. This source was detected by ML 6 times in the $q$-band, at different magnitudes between 19.84 and 20.11 in the $q$-band. The light curve did not show any significant variation, with a maximum change of $\sim$0.3 magnitude over 01:51:16 hours.

\textit{GaiaX21-117403--} (RA=342.90325$^{\circ}$, Dec=--70.19925$^{\circ}$) this \textit{GaiaX} Alert was detected on 2021-09-04 23:03:02 UTC at a magnitude of 19.52~$\pm$~0.0 and ML detected it on 2021-09-04 22:58:24 UTC at a magnitude of 19.74~$\pm$~0.12 in the $q$-band. We identified this source as a candidate supernova, possibly associated with the galaxy LEDA 274952 located at an offset of 4.4$\arcsec$ from the transient. It was detected by ML twice within 00:05:22 hours at different magnitude measurements of 19.74~$\pm$~0.11 and 19.83~$\pm$~0.12.

\textit{GaiaX21-118543--} (RA=306.25676$^{\circ}$, Dec=--64.11817$^{\circ}$) was detected by \textit{GaiaX} on 2021-09-08 21:31:41 UTC at a magnitude of 19.87~$\pm$~0.02 and by ML on 2021-09-08 21:28:32 UTC at a magnitude of 19.63~$\pm$~0.13 in the $q$-band. \textit{GaiaX} detected it 3 times, whereas ML detected it 5 times at different magnitudes (19.56-19.82). The source magnitude varies over a range of 0.03 magnitudes during the 1:48:39 hours that ML observations of this source took place.

\textit{GaiaX21-118564--} (RA=303.76967$^{\circ}$, Dec=--47.77041$^{\circ}$) On 2021-09-08 23:35:51 UTC, \textit{GaiaX} reported this alert at a magnitude of 19.52~$\pm$~0.02. It was also detected by ML on 2021-09-08 23:30:00 UTC at a magnitude of 19.46~$\pm$~0.14 in the $q$-band.

\textit{GaiaX21-118592--} (RA=339.46975$^{\circ}$, Dec=--88.21035$^{\circ}$) \textit{GaiaX} reported this alert on 2021-09-09 03:05:53 UTC by ML at a magnitude of 19.70~$\pm$~0.13 in the $q$-band. 


\subsection{List \#2: GaiaX: Yes, ML: No}
List \#2 involved investigating why \textit{GaiaX} triggered an alert when ML did not detect a transient. There could be several reasons for this, such as the transient being too faint to be detected by ML for instance due to bad observing conditions at ML. We compared the magnitude detection limits to check if the ML detection limit is deeper than the \textit{GaiaX} limit, implying that ML should have detected the transient. If so, it suggests that the \textit{GaiaX} Alert is spurious. There were 21 such detections (see Table \ref{tab:list2}). We also checked for any \textit{GaiaX} alerts in which the ML detection limit was not deep enough to observe the candidate transient and we found no such cases.

\subsection{List \#3: GaiaX: No, ML: Yes}
We uncover that there are 47 ML candidate transient detections in the last list where ML detected a transient that in principle, \textit{GaiaX} could have also detected but did not (Table~\ref{tab:list3}). There could be several reasons for this, which we discuss below. 

\begin{figure} 
\centering 
\includegraphics[scale=0.34]{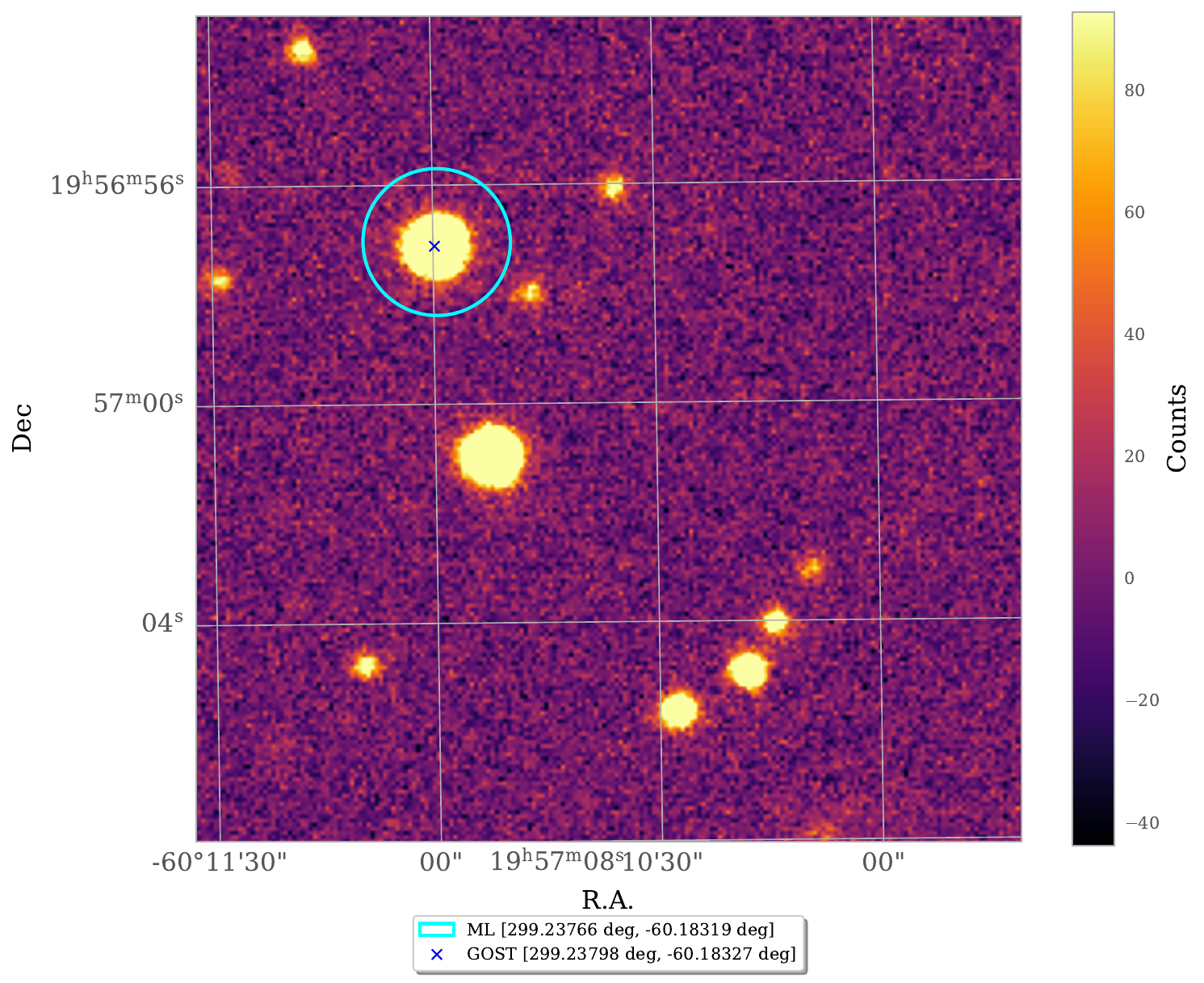}
\caption{ML reported a detection at RA = 299.23766$\mathrm{^\circ}$, Dec = --60.18318$\mathrm{^\circ}$ on 2021-09-09 21:31:04 UTC at a magnitude of 18.24$\pm$0.08 in the $q$-band, indicated by the cyan circle with a radius of 10$\arcsec$ (the astrometric accuracy of ML is 0.01$\arcsec$, but for a clearer plot, we chose a radius of 10$\arcsec$ to circle the detection). According to GOST, {\it Gaia} was at the position RA = 299.23798$\mathrm{^\circ}$, Dec = --60.18326$\mathrm{^\circ}$ (denoted with the blue cross) within 10 minutes of the ML detection, but there was no \textit{GaiaX} alert since the candidate transient is near a bright source.}
\label{fig:list3_bright}
\end{figure}

\begin{figure} 
\centering
\includegraphics[scale=0.34]{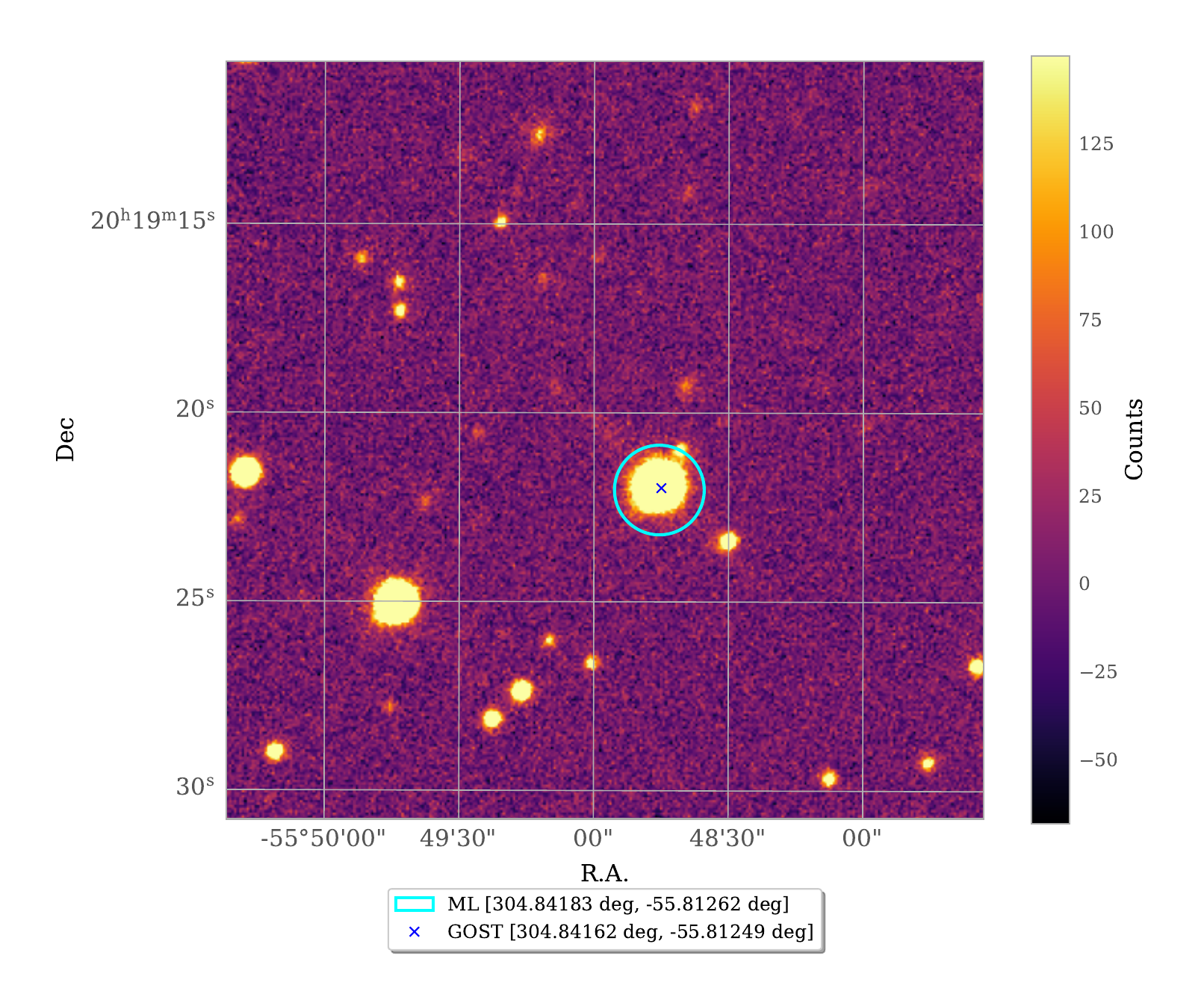}
\caption{On 2021-09-08 17:20:33 UTC, ML detected a candidate transient at RA = 304.84183$\mathrm{^\circ}$, Dec = --55.81261$\mathrm{^\circ}$ at a magnitude of 15.30$\pm$0.02 in the $q$-band, indicated by the cyan circle with a radius of 10$\arcsec$ (the astrometric accuracy of ML is 0.1$\arcsec$, but for a clearer plot, we chose a radius of 10$\arcsec$ to circle the detection). According to GOST, {\it Gaia} was at the position RA = 304.84161$\mathrm{^\circ}$, Dec = --55.81248$\mathrm{^\circ}$ (denoted with the blue cross) within 10 minutes of the ML detection. The source was reported in the GDR3 catalogue \citep{gaia_dr3} at an offset of 0.7$\arcsec$, catalogued with the ID 6471863812052661632, and therefore was not reported by the \textit{GaiaX} algorithm. This ML transient was also previously detected 6 times, with the first detection in 2019. The ML light curve of this object shows slight variability over this period, with detected magnitudes between 14.02 and 15.96. This candidate transient detection was therefore potentially caused by a variable star.}
\label{fig:list3_gdr2}
\end{figure}

The first possibility is that the source is near a bright star (e.g., Figure~\ref{fig:list3_bright}), which implies that the \textit{GaiaX} algorithm has removed it owing to its proximity to said bright source. There were only 3 such ML candidate transient detections. Second, the source was already reported and classified in the {\it Gaia} Data Releases (\citealt{brown_dr1, gdr2, gaia_dr3}) and was subsequently removed from the \textit{GaiaX} pipeline (e.g., Figure~\ref{fig:list3_gdr2}). There were 20 such candidate transient detections that had a matching source within 1$\arcsec$ in GDR3. Third, there is a possibility that the ML candidate transient detections are spurious in nature. On occasion, ML detects a false-positive transient which is incorrectly assigned a class\_real value of $\geq$~0.7 by the machine-learning algorithms. This occurs due to the inadequate representation of real and bogus transients in the machine learning training sets \citep{meercrab}. In this case, there were 24 spurious ML detections which were further confirmed by visual inspection of the images. Fourth, the candidate transient is an asteroid detection by ML. Many asteroids are detected by ML and while the known asteroids are accounted for in the ML and GaiaX transient pipeline, a new asteroid could appear to be detected as a candidate transient. We cross-checked the 47 ML candidate transients with the MPChecker (Minor Planet Checker) and found no known asteroids.

\begin{figure}
\centering
\includegraphics[width=\linewidth]{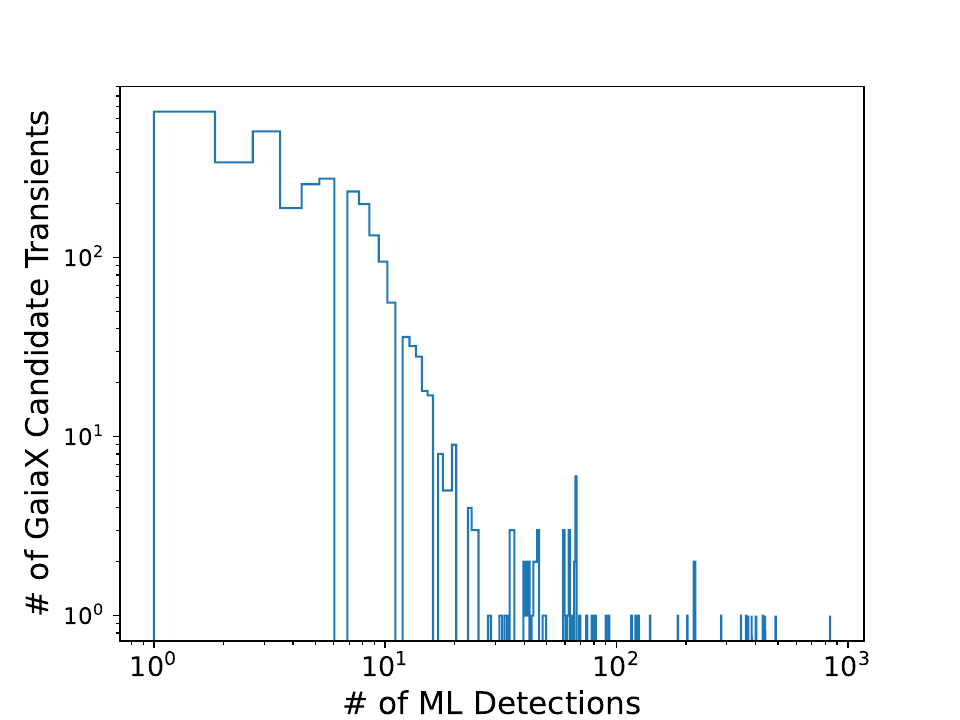}
\caption{Distribution of the number of ML detections for the 3181 \textit{GaiaX} candidate transients with at least one ML detection. The ML forced photometry routine returned 3181 unique \textit{GaiaX} candidate transients with corresponding significant ML detections. The highest number of corresponding ML detections is 804 for GaiaX21-115853, which is a variable star (see light curve in Figure~\ref{fig:variable}).}
\label{fig:detnum}
\end{figure}

\begin{figure*} 
    \begin{minipage}[b]{0.5\linewidth} 
        \centering
        \includegraphics[width=\textwidth]{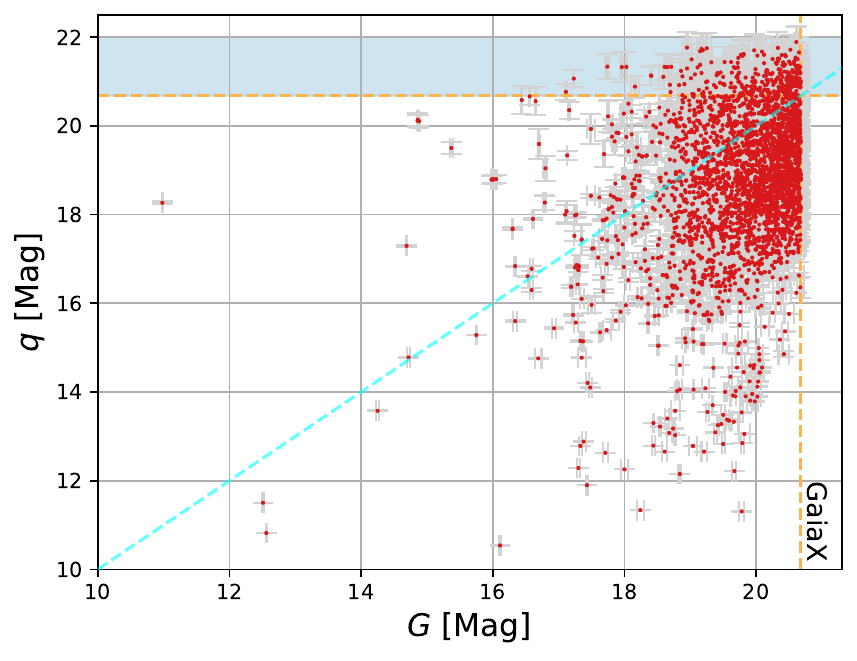}
    \end{minipage}%
    \begin{minipage}[b]{0.5\linewidth} 
        \centering
        \includegraphics[width=\textwidth]{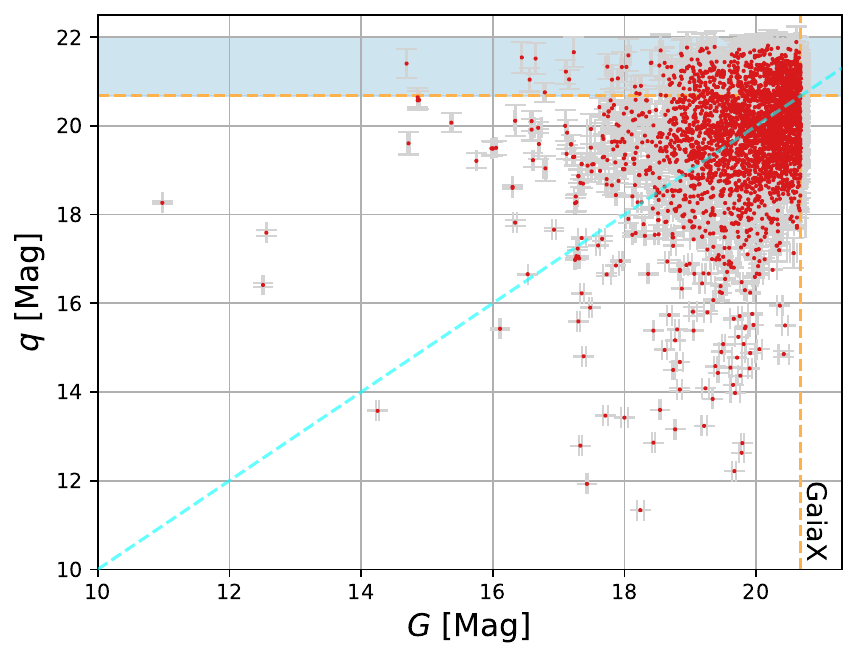}
    \end{minipage}
    \caption{\textit{GaiaX} G-band magnitude vs. ML $q$-band magnitude of {\it GaiaX} transient candidates. The \textbf{ML} forced photometry routine is run on the 7572 unique \textit{GaiaX} alerts to find any matching ML sources at the exact position of the \textit{GaiaX} alert, i.e. at the corresponding pixel of the ML CCD. This returned 3181 unique \textit{GaiaX} candidate transients for which ML had observations, denoted by the red data points with the error bars in gray. The orange dashed line refers to the limiting magnitude of \textit{GaiaX}, where in the y-axis, we assumed $q$ -- G = 0. The blue shaded region contains sources that are detected by ML at a magnitude below the {\it Gaia} DR3 limiting magnitude and therefore these sources may well have had a flare during which they were detected by {\it GaiaX}. Likewise, sources above the $q$ -- G = 0 line (the dashed cyan line) brightened in magnitude between the \textit{GaiaX} and ML detections, and sources below the line dimmed, barring any $q-$G colour effects. \textit{Left Panel:} For every \textit{GaiaX} candidate transient, the brightest corresponding ML detection was plotted. \textit{Right Panel:} For every \textit{GaiaX} candidate transient, the faintest corresponding ML detection was plotted. See Table~\ref{tab:numbers_fig8} for the corresponding numbers of each case and Section~\ref{forcedphot} for more details.}
    \label{fig:G_vs_q}
\end{figure*}

\begin{figure*}
    \centering
    \begin{subfigure}[b]{0.45\linewidth}
        \centering
        \includegraphics[width=\textwidth]{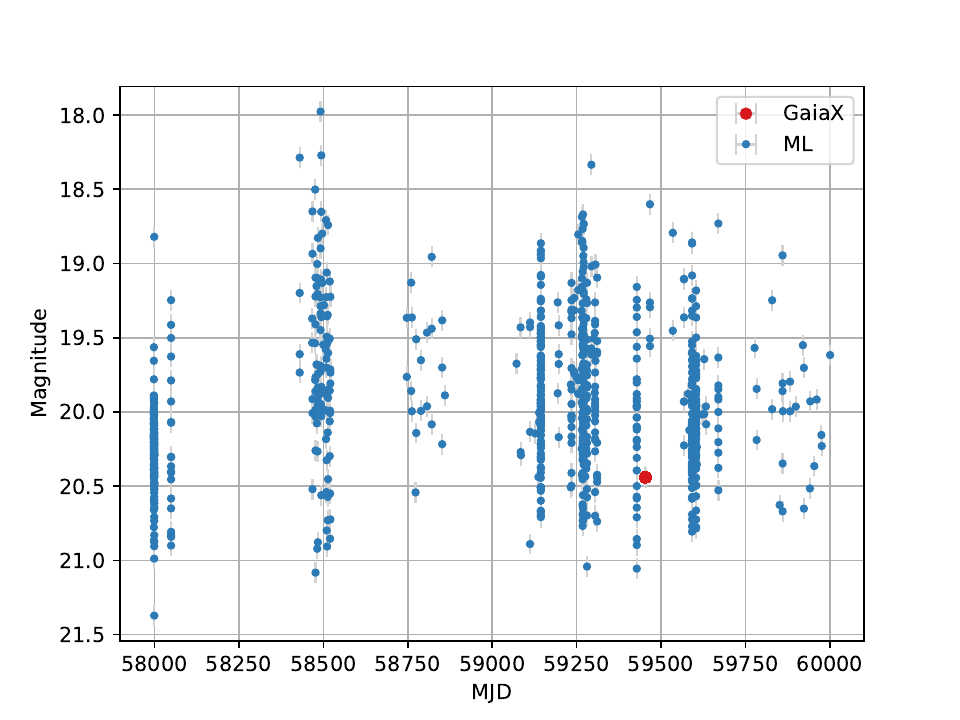}
        \caption{GaiaX21-115853}
        \label{fig:variable}
    \end{subfigure}
    \hfill
    \begin{subfigure}[b]{0.45\linewidth}
        \centering
        \includegraphics[width=\textwidth]{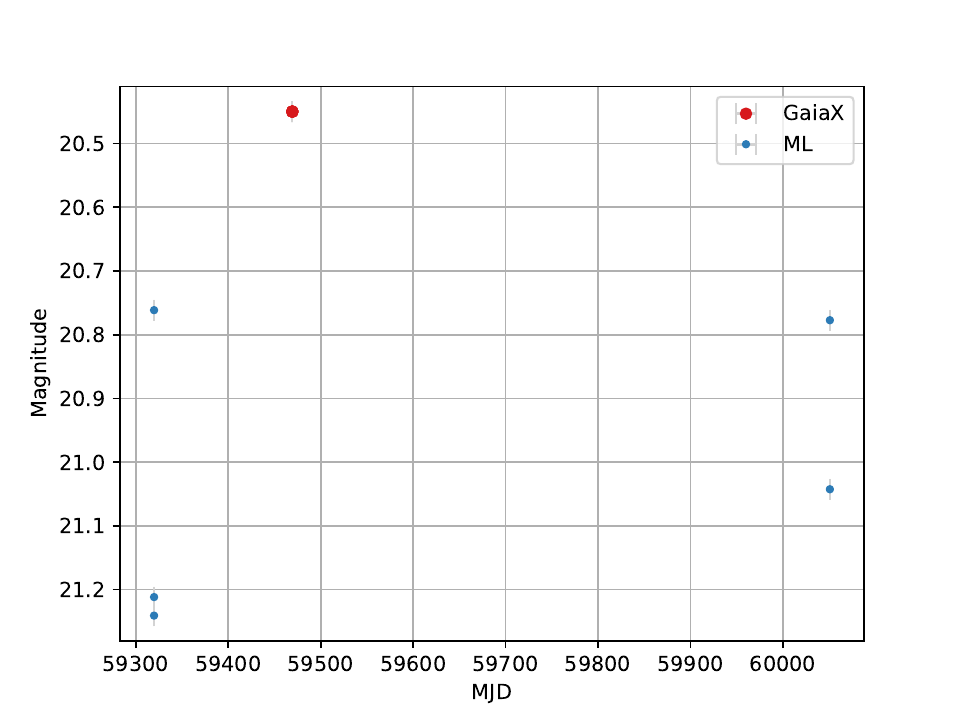}
        \caption{GaiaX21-119276}
        \label{fig:toofaint}
    \end{subfigure}

    \vspace{0.5cm}
    
    \begin{subfigure}[b]{0.45\linewidth}
        \centering
        \includegraphics[width=\textwidth]{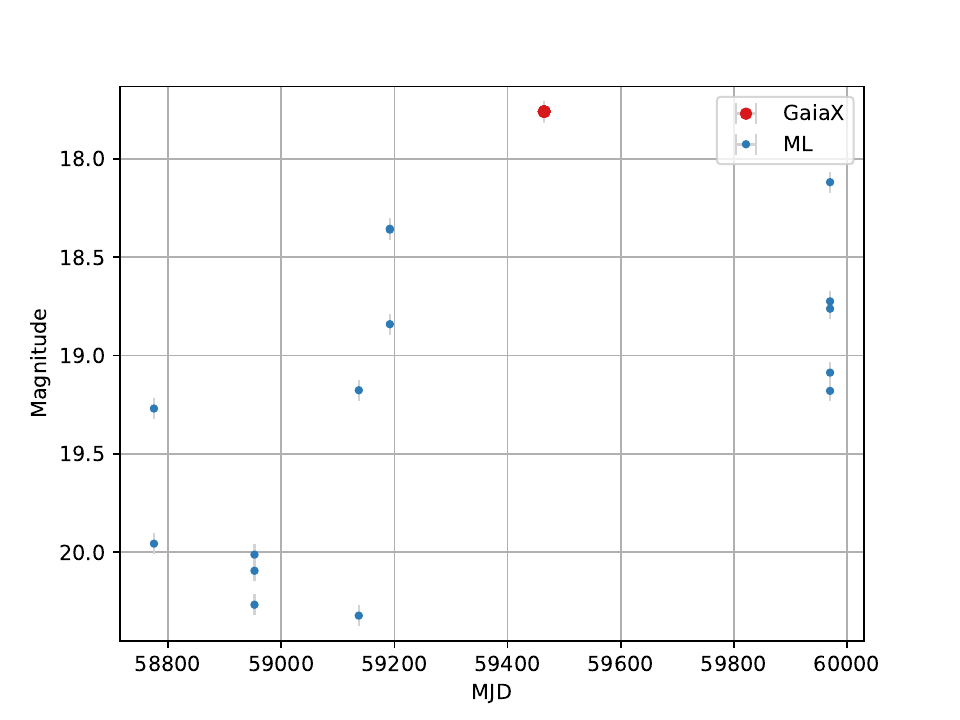}
        \caption{GaiaX21-118137}
        \label{fig:brightened}
    \end{subfigure}
    \hfill
    \begin{subfigure}[b]{0.45\linewidth}
        \centering
        \includegraphics[width=\textwidth]{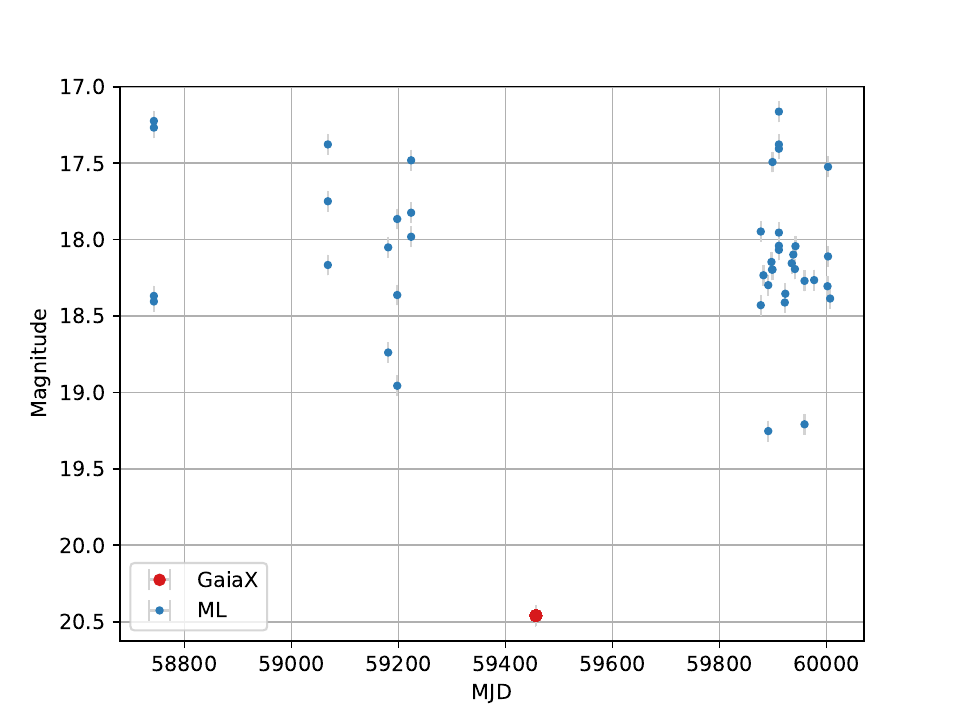}
        \caption{GaiaX21-116085}
        \label{fig:dimmed}
    \end{subfigure}
    
    \caption{Light curve examples of \textit{GaiaX} candidate transients (in red) that have ML detections at the same position (in blue), based on the forced photometry results shown in Figure~\ref{fig:G_vs_q} (see Section~\ref{forcedphot} for more details): \textit{(a)} The light curve of a variable star \citep{varstar}, detected as a candidate transient with the ID \textit{GaiaX21-115853} by \textit{GaiaX}. On further analysis of the light curve, the variable star appears to be varying sinusoidally over a period of approximately 2 months. \textit{(b)} The light curve of a candidate eclipsing binary \citep{eclipbinary} which was previously too faint to be detected by \textit{Gaia} but was detected by \textit{GaiaX} at a magnitude of 20.45~$\pm$~0.04. \textit{(c)} The light curve of a supernova candidate \citep{2021SNcand} (SN 2021xjj) detected by \textit{GaiaX} as a transient, with the ML detections likely corresponding to the diffuse light from the host galaxy at the same position. \textit{(d)} The light curve of the galaxy LEDA 435505 detected with different flux values of its diffuse light by ML due to the variation in seeing conditions.}
    \label{fig:lightcurves}
\end{figure*}

\begin{table*}
\caption{\label{tab:numbers_forcedphot} We ran the ML forced photometry routine on all the 212381 ML images (up to May 2023) for the 7572 unique \textit{GaiaX} candidate transients, which we further filtered in a series of steps based on set conditions that are summarized below.}
\centering
\begin{tabular}{|M{8cm}||M{3cm}|M{3cm}|}
\hline
\textbf{Steps} & \textbf{\# of ML images} & \textbf{\# of Corresponding \textit{GaiaX} Alerts} \\
\hline
ML forced photometry results  & 148980 & 7572 \\
\hline
Remove all the entries flagged as "bad" and all the null detections & 80075 & -- \\
\hline 
Filter based on condition SNR\_OPT > 3 & 25560 & 3181 \\
\hline
Filter based on condition SNR\_OPT < 3 & 54515 & 3622 \\
\hline 
Check for \textit{GaiaX} candidate transients with no matching ML detections & -- & 769 \\
\hline
\end{tabular}
\end{table*}

\subsection{\textit{GaiaX} candidate transients and ML forced photometry} \label{forcedphot}
In addition to the lists mentioned above, we try to find out how many of the \textit{GaiaX} alerts issued during the experiment are real sources. To do this, we ran the ML forced photometry routine on all the 212381 ML images (up to May 2023) for the 7572 unique \textit{GaiaX} candidate transients individually to check for detections at the exact position, i.e. at the corresponding pixel on the ML CCD. This returned 148980 ML values, which we further filtered. We first removed all the entries flagged as "bad" by the detection pipeline, and all the null detections, which resulted in 80075 ML magnitudes. The null detections refer to detections in which the object flux is negative. We further filtered this based on the optimal SNR or SNR\_OPT values, as described in Section~\ref{sec:real}. This resulted in 25560 ML detections with SNR\_OPT $>3$, and 54515 entries with SNR\_OPT $<3$ which we treat as upper limits. The 25560 significant ML detections correspond to 3181 unique \textit{GaiaX} candidate transients. Out of the remaining 4391 unique \textit{GaiaX} candidate transients, 3622 candidates had non-significant ML images at the same position and 769 \textit{GaiaX} candidates had no corresponding ML detections. The distribution of the number of ML detections for each unique \textit{GaiaX} candidate transient can be seen in Figure~\ref{fig:detnum}. 

In Figure~\ref{fig:G_vs_q}, we compare and plot one significant ML detection for every unique \textit{GaiaX} candidate transient. For every unique \textit{GaiaX} candidate transient, we consider the brightest ML detection (\textit{Left Panel}) and the faintest ML detection (\textit{Right Panel}). We discuss two different cases below, with examples, and in Table~\ref{tab:numbers_fig8}, we summarize the corresponding numbers of each case.

\begin{table*}
\caption{\label{tab:numbers_fig8} From the ML forced photometry results, we have 3181 unique \textit{GaiaX} candidate transients with corresponding significant ML detections. In Figure~\ref{fig:G_vs_q}, we plot one significant ML detection for every unique \textit{GaiaX} candidate transient, and consider the brightest ML detections \textit{(Left Panel)} and the faintest ML detection \textit{(Right Panel)}. Below, we mention the corresponding numbers for the two different cases.}
\centering
\begin{tabular}{|M{6cm}||M{3cm}|M{3cm}|}
\hline
\multicolumn{3}{|c|}{\textbf{\textit{GaiaX} Alerts Corresponding to ML Detections}} \\
\hline 
 & \textbf{Brightest ML Detections} & \textbf{Faintest ML Detections} \\
\hline
Brightened sources i.e. $q$ [Mag] $<$ G [mag] & 1203 & 1664 \\
$q$ [Mag] $>$ 20.68 & 485 & 736 \\
\hline 
Dimmed sources i.e. $q$ [Mag] $>$ G [mag] & 1978 & 1517 \\
\hline
\end{tabular}
\end{table*}

\begin{enumerate}

    \item \textbf{Brightened sources i.e. $q$ [Mag] $<$ G [mag]:} Some of the \textit{GaiaX} candidate transients brightened in magnitude, between the \textit{GaiaX} and ML detection times, and these sources lie above the $q$ -- G = 0 dashed cyan line. For example, in Figure~\ref{fig:brightened}, the \textit{GaiaX} detection is a candidate supernova detection that was detected by ASAS-SN on 2021-08-31 08:52 UTC \citep{2021SNcand} (SN 2021xjj) and was detected by \textit{GaiaX} on 2021-09-07 16:12 UTC at an offset of 1.4$\arcsec$ and at a magnitude of 17.76~$\pm$~0.01. The previous ML detections at the same position are due to diffuse light from the host galaxy at the \textit{GaiaX} transient position. The forced photometry routine measures the  flux at the {\it GaiaX} position in the associated ML images, however there could be different flux measurements due to variations in the seeing. When we consider the brightest ML detections, there are 1203 brightened sources. Similarly, when we consider the faintest ML detections, there are 1664 brightened sources.
    
    A number of ML detections have magnitudes greater than the limiting magnitude of \textit{GaiaX} i.e. $q$ [Mag] $>$ 20.68. The limiting magnitude of \textit{GaiaX} is 20.68, whereas for ML, it strongly depends on the (seeing) conditions at Sutherland, but is approximately 21 magnitude in most cases. In certain cases, the source brightened and was detected by \textit{GaiaX} as a candidate transient but was previously too faint to be detected by \textit{Gaia}. For example, in Figure~\ref{fig:toofaint}, we see the light curve of a candidate eclipsing binary \citep{eclipbinary} which was previously too faint to be detected by \textit{Gaia} but \textit{GaiaX} detected it at a magnitude of 20.45~$\pm$~0.04. When we consider the brightest ML detections, there are 485 such faint and significantly detected ML sources and similarly, there are 736 ML sources when we consider the faintest ML detections.

    \item \textbf{Dimmed sources i.e. $q$ [Mag] $>$ G [mag]:} The detections that lie below the $q$ -- G = 0 dashed cyan line are sources that dimmed in magnitude between the \textit{GaiaX} and ML detections. This could be attributed to the detection of a variable star, such as in Figure~\ref{fig:variable} \citep{varstar} which \textit{GaiaX} detected at a magnitude of 20.44~$\pm$~0.08. On further analysis of the light curve, the variable star appears to be varying sinusoidally over a period of approximately 2 months. It could also be a detection of an extended source such as a galaxy, for which ML measured different flux measurements of the diffuse light from the galaxy due to variation in the seeing conditions, at the exact position of the \textit{GaiaX} candidate transient position. Figure~\ref{fig:dimmed} is an example light curve of an extended source, the galaxy LEDA 435505, that ML detected with different flux values. at an offset of 0.5$\arcsec$ from the \textit{GaiaX} candidate transient position, which was detected at a magnitude of 20.46~$\pm$~0.08 by \textit{GaiaX}. When we consider the brightest ML detections, there are 1978 dimmed sources and when we consider the faintest ML detections, there are 1517 dimmed sources.
\end{enumerate}

\section{Final Remarks}
\label{sec:remarks}
Ahead of the current observing run of the LIGO-Virgo-KAGRA (LVK) collaboration, O4, we ran an experiment with \textit{GaiaX} and ML to test the new \textit{GaiaX} detection pipeline, which is specifically designed to detect the optical counterparts of GW events. \textit{GaiaX} will run independent of the GW event triggers, but the alerts can coincide with a GW event sky localization area during O4. With these public \textit{GaiaX} alerts, users can check and crossmatch their potential candidates, and also use catalogues such as the Transient Name Server (TNS) and the Minor Planet Center (MPC).

The experiment consisted of ML observing the same region of the sky as \textit{Gaia} within 10 minutes, whenever the visibilities overlapped and the seeing conditions were suitable, during the period of two weeks centred on New Moon over about two months. ML could successfully do this for $\sim$35.7\% of the total time. This resulted in 7572 unique \textit{GaiaX} alerts of candidate transients, which we investigate further. During the time when the two telescopes were indeed observing the same position on the sky within 10 minutes, we found 27 unique \textit{GaiaX} candidate transients that have a corresponding ML detection. Out of this, we found 6 real transients (Figure~\ref{fig:real_transients}) while the remaining 21 are likely spurious. For the 47 ML candidate transient detections that were detected when \textit{GaiaX} and ML were observing the same region of the sky within 10 minutes, but there was no \textit{GaiaX} alert, we see examples of sources near bright sources (Figure~\ref{fig:list3_bright}) or that were previously reported in the \textit{Gaia} data releases (Figure~\ref{fig:list3_gdr2}) and thus, were filtered out in the \textit{GaiaX} detection pipeline. 

Additionally, we ran the ML forced photometry routine on all the unique \textit{GaiaX} candidate transients to find matching ML sources that were detected before or after the {\it GaiaX} detection. We found 3181 unique \textit{GaiaX} candidate transients with a corresponding significant ML detection (Figure~\ref{fig:G_vs_q}). In many cases, these were detections of sources that had a flare in their magnitude and was detected as a transient by \textit{GaiaX}. There were also detections of sources that were too faint to be detected and appear in GDR3 which is taken as the input catalogue for \textit{GaiaX}, and hence appeared as a detection by \textit{GaiaX}. Many of these sources consist of several types of objects such as variable stars, galaxies measured with different flux values (due to seeing), and eclipsing binaries. Some of these candidate transients are also due to real non-repeating transient phenomena, like supernovae (Figure~\ref{fig:lightcurves}). An outcome of our work is that it is important to check whether the \textit{GaiaX} alerts can be due to a faint source that flared up by comparing the \textit{GaiaX} transient position with previous or later observations that covered the position of the candidate transient. 

During the previous observing run of the LIGO-Virgo Collaboration (LVC), O3, the possibility of \textit{Gaia} contributing to the search for optical counterparts of GW events was considered and explored with the existing GSA pipeline and the LVC GW event skymaps. This resulted in 12 GCN circulars being published. In the ongoing observing run, O4, the enhanced detection pipeline \textit{GaiaX}\footnote{\label{GaiaX}\href{http://gsaweb.ast.cam.ac.uk/alerts/gaiax/}{http://gsaweb.ast.cam.ac.uk/alerts/gaiax/}} is switched on so we expect to detect more possible GW counterparts.

\section{Acknowledgements}

SB would like to thank Ashley Chrimes for his helpful inputs during discussions related to this project. SB acknowledges studentship support from the Dutch Research Council (NWO) under the project number 680.92.18.02. ZKR acknowledges funding from the Netherlands Research School for Astronomy (NOVA). This work has made use of data from the European Space Agency (ESA) mission {\it Gaia} (\url{https://www.cosmos.esa.int/gaia}), processed by the {\it Gaia} Data Processing and Analysis Consortium (DPAC, \url{https://www.cosmos.esa.int/web/gaia/dpac/consortium}). Funding for the DPAC has been provided by national institutions, in particular the institutions participating in the {\it Gaia} Multilateral Agreement. This research has made use of the SIMBAD database, operated at CDS, Strasbourg, France.  This research has made use of data and/or services provided by the International Astronomical Union's Minor Planet Center. This work made use of Astropy:\footnote{\href{http://www.astropy.org}{http://www.astropy.org}} a community-developed core Python package and an ecosystem of tools and resources for astronomy \citep{astropy:2013, astropy:2018, astropy:2022}.

\section{Data Availability}
The \textit{GaiaX} Alerts data from the test phase can be found here: \url{http://gsaweb.ast.cam.ac.uk/alerts/gaiaxtest/}. All other data will be made available in a reproduction package uploaded to Github. 
 



\bibliographystyle{mnras}
\bibliography{example} 




\appendix

\newpage
\section{Detection Details}

\begin{table*}
\caption{\label{tab:list1}List \#1}
\centering
\begin{tabular}{|M{2.5cm}|M{2cm}|M{2.8cm}|M{1.7cm}|M{2.8cm}|M{1.7cm}|M{0.8cm}|}
\hline
\textbf{GaiaX ID} & \textbf{MJD} & \textbf{GaiaX (RA, Dec) [deg]} & \textbf{GaiaX G Mag} & \textbf{ML (RA, Dec) [deg]} & \textbf{ML Mag} & \textbf{Filter} \\
\hline
GaiaX21-118536 & 59465.9518 &  337.55000, --86.31721 &  19.89~$\pm$~0.04 & 337.54916, -86.31731 & 20.02~$\pm$~0.16 & $q$ \\
\hline
GaiaX21-118560 & 59465.8834 & 307.08512, --74.79555 & 20.13~$\pm$~0.02 & 307.08502, -74.79550 & 20.11~$\pm$~0.18  & $q$ \\
\hline 
GaiaX21-117403 & 59461.8824 & 342.90341, --70.19923 & 19.52~$\pm$~0.0 & 342.90313, -70.19910 & 19.74~$\pm$~0.12 & $q$ \\
\hline 
GaiaX21-118543 & 59465.8929 & 306.25655, --64.11816 & 19.87~$\pm$~0.02 & 306.25664, -64.11822 & 19.63~$\pm$~0.13 & $q$ \\
\hline
GaiaX21-118564 & 59465.9791 & 303.76981, --47.77047 & 19.52~$\pm$~0.02 & 303.76963, -47.77040 & 19.46~$\pm$~0.14 & $q$ \\
\hline
GaiaX21-118592 & 59466.1290 & 339.46995, --88.21044 & 19.81~$\pm$~0.01 & 339.47095, -88.21042 & 19.70~$\pm$~0.13 & $q$ \\
\hline
\end{tabular}
\end{table*}

\begin{table*}
\caption{\label{tab:realtransient}List \#1 ML light curve information (order corresponding to Figure~\ref{fig:real_transients})}
\centering
\begin{tabular}{|M{2cm}|M{2cm}|M{0.7cm}|M{2.8cm}|M{1cm}|M{1.1cm}|M{1cm}|M{1cm}|M{1cm}|}
\hline
\textbf{GaiaX ID} & \textbf{MJD} & \textbf{Filter} & \textbf{RA, Dec [deg]} & \textbf{SNR ZOGY} & \textbf{Mag} & \textbf{Mag Error} & \textbf{Flux [$\mu Jy$]} & \textbf{Flux Error [$\mu Jy$]} \\
\hline
GaiaX21-118536 & 59465.95184 & $q$ & 337.54984, --86.31725 & 6.770 & 20.02 & 0.16 & 35.52 & 5.26 \\
\hline 
GaiaX21-118560 & 59465.88342 & $q$ & 307.08475, --74.79558 & 6.230 & 20.11 & 0.18 & 32.75 & 5.38 \\
 & 59465.88434 & $q$ & 307.08435, --74.79552 & 6.730 & 20.05 & 0.19 & 34.82 & 5.39 \\
 & 59465.88526 & $q$ & 307.08498, --74.79560 & 6.973 & 20.03 & 0.16 & 35.30 & 5.28 \\
 & 59465.88618 & $q$ & 307.08514, --74.79546 & 7.424 & 19.91 & 0.15 & 39.30 & 5.34 \\
 & 59465.88708 & $q$ & 307.08495, --74.79562 & 6.668 & 20.06 & 0.17 & 34.29 & 5.24 \\
 & 59465.96069 & $q$ & 307.08514, --74.79565 & 7.118 & 19.85 & 0.16 & 41.82 & 6.12 \\
\hline
GaiaX21-117403 & 59461.88246 & $q$ & 342.90325, --70.19925 & 9.300 & 19.74 & 0.12 & 46.09 & 4.93 \\
& 59461.95723 & $q$ & 342.90350, --70.19924 & 8.735 & 19.83 & 0.12 & 42.42 & 4.71 \\
\hline
GaiaX21-118543 & 59465.89299 & $q$ & 306.25676, --64.11817 & 8.493 & 19.63 & 0.13 & 51.21 & 5.94 \\
 & 59465.89391 & $q$ & 306.25669, --64.11809 & 7.967 & 19.77 & 0.13 & 44.84 & 5.55 \\
 & 59465.89482 & $q$ & 306.25704, --64.11811 & 8.016 & 19.70 & 0.13 & 47.67 & 5.84 \\
 & 59465.96751 & $q$ & 306.25653, --64.11818 & 6.781 & 19.82 & 0.16 & 42.51 & 6.13 \\
 & 59465.96844 & $q$ & 306.25654, --64.11815 & 6.863 & 19.57 & 0.16 & 54.15 & 7.89 \\
\hline
GaiaX21-118564 & 59465.97917 & $q$ & 303.76967, --47.77041 & 7.762 & 19.46 & 0.14 & 59.77 & 7.73 \\
\hline
GaiaX21-118592 & 59466.1291 & $q$ & 339.46975, --88.21035 & 8.18 & 19.71 & 0.13 & 47.67 & 5.57 \\
\hline
\end{tabular}
\end{table*}

\begin{table*}
\caption{\label{tab:list2}List \#2 (ordered by MJD)}
\centering
\begin{tabular}{|M{2cm}|M{2cm}|M{2.8cm}|M{2cm}|M{2cm}|}
\hline
\textbf{GaiaX ID} & \textbf{MJD} & \textbf{RA, Dec [deg]} & \textbf{GMag} & \textbf{GMag Err} \\
\hline
GaiaX21-116833 & 59459.98260 & 308.44874, --43.49446 & 20.39 & 0.02 \\
\hline
GaiaX21-116834 & 59459.98427 & 306.84985, --41.46077 & 20.32 & 0.03 \\
\hline
GaiaX21-117307 & 59461.88261 & 356.70694, --72.81483 & 20.38 & 0.04 \\
\hline 
GaiaX21-117308 & 59461.88658 & 343.66312, --69.61603 & 19.76 & 0.12 \\
\hline
GaiaX21-117311 & 59461.91003 & 308.32648, --43.16348 & 20.19 & 0.02 \\
\hline 
GaiaX21-117313 & 59461.91172 & 306.86682, --41.03549 & 20.29 & 0.02 \\
\hline 
GaiaX21-117314 & 59461.91247 & 306.22101, --40.06524 & 20.32 & 0.05 \\
\hline 
GaiaX21-117401 & 59461.91248 & 306.20462, --40.06293 & 20.48 & 0.03 \\
\hline 
GaiaX21-117405 & 59461.98572 & 306.86832, --41.01888 & 20.28 & 0.04 \\
\hline 
GaiaX21-117406 & 59461.98648 & 306.21504, --40.06340 & 20.24 & 0.05 \\
\hline 
GaiaX21-117322 & 59461.98648 & 306.19586, --40.06471 & 20.49 & 0.02 \\
\hline 
GaiaX21-118503 & 59465.73169 & 304.77269, --49.67635 & 20.32 & 0.05 \\
\hline 
GaiaX21-118505 & 59465.73416 & 303.70349, --46.30773 & 20.24 & 0.04 \\
\hline
GaiaX21-118538 & 59465.88844 & 310.37705, --75.35009 & 20.09 & 0.08 \\
\hline
GaiaX21-118365 & 59465.90306 & 304.46114, --56.04829 & 18.98 & 0.04 \\
\hline
GaiaX21-118366 & 59465.90515 & 304.07335, --53.25542 & 19.08 & 0.03 \\
\hline 
GaiaX21-118545 & 59465.91033 & 303.70521, --46.27803 & 20.18 & 0.07 \\
\hline 
GaiaX21-118726 & 59466.73423 & 301.17305, --46.83247 &  20.39 & 0.02 \\
\hline 
GaiaX21-118759 & 59466.90467 & 300.37367, --54.45753 & 19.17 & 0.02 \\
\hline 
GaiaX21-118760 & 59466.90517 & 300.90044, --53.83301 & 19.78 & 0.04 \\
\hline
\end{tabular}
\end{table*}

\begin{table*}
\caption{\label{tab:list3}List \#3 (ordered by MJD)}
\centering
\begin{tabular}{|M{2cm}|M{2cm}|M{2.8cm}|M{2cm}|M{2cm}|}
\hline
\textbf{ML ID} & \textbf{MJD} & \textbf{RA, Dec [deg]} & \textbf{Mag} & \textbf{Mag Err} \\
\hline
77818360 & 59461.88128 & 357.97071, -72.35282 & 19.24 & 0.08 \\ 
\hline
77823358 & 59461.90859 & 306.15683, -37.38415 & 19.03 & 0.06 \\
\hline
77824035 & 59461.95722 & 346.78395, -70.76889 & 18.59 & 0.04 \\
\hline
77824234 & 59461.95827 & 341.20218, -68.92360 & 17.26 & 0.02 \\
\hline
77824483 & 59461.95931 & 336.67696, -67.62492 & 13.79 & 0.01 \\
\hline
77843960 & 59461.96319 & 328.70481, -64.61370 & 16.05 & 0.01 \\
\hline
77825961 & 59461.96699 & 322.61730, -59.71068 & 14.68 & 0.00 \\
\hline
77844325 & 59461.97074 & 317.69323, -54.40289 & 15.12 & 0.01 \\
\hline
77844572 & 59461.97177 & 315.35892, -52.39046 & 17.38 & 0.05 \\
\hline
77834515 & 59462.12300 & 31.36140, -74.81026 & 12.49 & 0.00 \\
\hline
77817907 & 59462.12812 & 0.45583, -73.96224 & 16.41 & 0.01 \\
\hline
77839150 & 59462.13101 & 352.79982, -72.56973 & 17.79 & 0.08 \\
\hline
77859565 & 59465.71963 & 306.85583, -61.41103 & 13.39 & 0.01 \\
\hline
77860600 & 59465.72066 & 307.57759, -59.78251 & 18.33 & 0.04 \\
\hline
77860897 & 59465.72261 & 304.84183, -55.81261 & 15.30 & 0.02 \\
\hline
77862278 & 59465.72739 & 305.40980, -49.99619 & 15.97 & 0.01 \\
\hline
77879746 & 59465.87697 & 14.979310, -88.07231 & 13.05 & 0.00 \\
\hline
77880479 & 59465.87953 & 328.22204, -83.43748 & 17.95 & 0.03 \\
\hline
77887024 & 59465.88912 &3 10.79846, -70.87467 & 16.28 & 0.03 \\
\hline
77891881 & 59465.89106 & 309.32221, -67.40380 & 12.96 & 0.00 \\ 
\hline
77901068 & 59465.89481 & 307.73627, -64.51705 & 17.96 & 0.04 \\
\hline
77910466 & 59465.90152 & 305.16016, -53.49018 & 17.77 & 0.02 \\
\hline
77911036 & 59465.90346 & 305.80370, -52.104729 & 18.12 &0.06 \\ 
\hline
77911995 & 59465.90539 & 304.86301, -46.06155 & 12.34 & 0.01 \\
\hline
77919051 & 59465.95184 & 346.88316, -86.65516 & 17.88 & 0.05 \\
\hline
77919385 & 59465.95301 & 327.07057, -84.09406 & 17.52 & 0.05 \\
\hline
77920911 & 59465.95681 & 318.64075, -81.80236 & 17.15 & 0.04 \\
\hline
77921364 & 59465.95875 & 314.03302, -78.74489 & 17.40 & 0.05 \\
\hline
77922453 & 59465.95977 & 311.72972, -75.91257 & 12.28 & 0.00 \\
\hline
77924741 & 59465.96261 & 309.37881, -72.70382 & 13.12 & 0.01 \\
\hline
77926228 & 59465.96456 & 307.83618, -68.09910 & 17.25 & 0.05 \\
\hline
77929706 & 59465.96649 & 306.53886, -64.97557 & 14.44 & 0.02 \\
\hline
77932811 & 59465.97916 & 304.28488, -48.85948 & 18.61 & 0.07 \\
\hline
77946929 & 59466.12909 & 2.04691, -88.22453 & 17.83 & 0.06 \\
\hline
77947417 & 59466.72088 & 302.13884, -63.32795 & 17.01 & 0.06 \\
\hline
77947684 & 59466.72191 & 302.43440, -57.83077 & 13.58 & 0.01 \\
\hline
77950823 & 59466.73040 & 302.62282, -46.12839 & 16.66 & 0.03 \\
\hline
77955361 & 59466.87918 & 289.98533, -86.76007 & 17.81 & 0.06 \\
\hline
77956287 & 59466.88120 & 294.35898, -83.69677 & 17.17 & 0.03 \\
\hline
77956812 & 59466.88409 & 299.23010,-78.04095 & 13.19 & 0.01 \\
\hline
77957654 & 59466.88697 & 299.88442, -75.77201 & 16.98 & 0.06 \\
\hline
77957854 & 59466.88801 & 300.24347, -71.74227 & 13.07 & 0.01 \\
\hline
77959019 & 59466.89178 & 301.40145, -67.21155 & 18.26 & 0.05 \\
\hline
77959696 & 59466.89371 & 300.66087, -63.95414 &13.23 & 0.05 \\
\hline
77960767 & 59466.89657 & 299.23766, –60.18318 & 18.25 & 0.09 \\
\hline
77962782 & 59466.90127 & 301.68637, -54.82228 & 13.45 & 0.01 \\
\hline
77988054 & 59466.98092 & 301.97307, -48.90062 & 18.37 & 0.07 \\
\hline
\end{tabular}
\end{table*}


\bsp	
\label{lastpage}
\end{document}